\documentclass[epjH]{svjour}
\usepackage[most]{tcolorbox}
\usepackage{color,soul}
\usepackage{lipsum}
\usepackage{caption} 
\usepackage{graphicx}
\usepackage{hyperref}

\begin{document}
\title{The LHC Timeline: A Personal Recollection (1980-2012)\footnote{The text presented here has been revised by the authors based on the original oral history interview conducted by Luisa Bonolis and recorded in Rome, Italy, 1--3 March 2016.}
\subtitle{Oral History Interview}}

\author{
Luciano Maiani\inst{1}\fnmsep\thanks{\email{luciano.maiani@roma1.infn.it}}
\and Luisa Bonolis\inst{2}\fnmsep\thanks{\email{lbonolis@mpiwg-berlin.mpg.de, luisa.bonolis@roma1.infn.it}}
}
\institute{Dipartimento di Fisica and INFN, Piazzale A. Moro 5, 00185 Rome, Italy \and  Max Planck Institute for the History of Science, Boltzmannstra\ss e 22, 14195 Berlin, Germany}

\abstract{The object of this interview is the history of the Large Hadron Collider in the LEP tunnel at CERN,  from first ideas to the discovery of the Brout-Englert-Higgs boson, seen from the point of view of a member of CERN scientific committees, of the CERN Council and a former Director General of CERN in the years of machine construction.} 
%
\maketitle
\tableofcontents

\section{The $Sp{\bar p}S$ Collider and LEP}

L. B. \hspace{0.2 cm}
During the 1980s,  at the time when LEP, the Large Electron Positron collider, was being completed, you began to be deeply involved with CERN\dots\footnote{We are indebted to Prof. Dieter Haidt for  very informative exchange on the arguments discussed in this Section.}

\vskip 0.3 cm
L. M. \hspace{0.2 cm} Nicola Cabibbo had been in the Scientific Policy Committee (SPC) in the 1970s, when they had advised CERN to approve the construction of the proton-antiproton collider proposed by Carlo Rubbia and Simon van Der Meer to search for the Intermediate Vector Bosons. It was a difficult machine and the positive recommendation of the SPC was a far-sighted  decision.

The $p\bar p$-project of Rubbia, initiated in 1976, was based on the conclusion that with the unified theory of the electroweak interactions~\cite{Glashow1961}\cite{Weinberg1967}\cite{Salam1968} and the measurement of $\sin 2\theta$ from Gargamelle, the predicted mass of the charged intermediate vector boson $W$ was around $70$ GeV, requiring a proton-antiproton collider to reach such energies. However, at FermiLab, Bob Wilson was not interested in the project. 
Rubbia, after strong opposition, convinced finally L\'eon van Hove and John Adams of a $p\bar p$ collider, at the CERN SPS, based on the Stochastic Cooling project of Simon van der Meer. 

The S$p\bar p$S started operation in 1981, after only 5 years of construction. An immediate outcome was the appearance of narrow jets even in a hadron machine \cite{Banner1982}, a possibility doubted earlier by prominent physicists and a strong confirmation of the reality of confined quarks and gluons in the hadrons. The observation of $W$ and $Z$, with predicted masses and other properties followed in 1983, see e.g.~\cite{DiLella:2015yit}.

I was elected in the Scientific Policy Committee of CERN in 1984, after Nicola, when the Large Hadron Collider was  first considered in the SPC as the next CERN large facility.
 
  In those years,~
  we realised that the way to go to higher energies was with $p-p$ rather than $p-\bar{p}$ colliders or $e^+e^-$ colliders. At energies above LEP,  there are so many gluons in a proton or in an antiproton that new particle production will be dominated by gluon-gluon fusion. It was a great simplification: you would not need quark-antiquark annihilation, as was the case at the $Sp\bar p S$, and all the gymnastic around antiproton beam cooling could be avoided. Also, due to asymptotic freedom, we understood that a hadron machine could be almost as effective as an electron-positron machine to disentangle the basic particle reactions. So it didn't pay to go beyond LEP with electron-positron, the next machine could be proton-proton.
\vskip 0.2 cm

L. B. \hspace{0.2 cm} The years 1980s saw the realisation in Europe of two big high energy machines, HERA at DESY  and LEP at CERN. How was discussion and planning of two such large enterprises organised?

\vskip 0.2 cm
L. M. \hspace{0.2 cm} The scene was set by the rising of the Standard Model in the 1970s after the discoveries of neutral weak current neutrino events and of charm, see e.g. the reconstruction presented in \cite{Maiani2017}. 

New particles were expected to be observed at high energy ($W$ and $Z$, new quark and lepton flavours) and the theoretical progress after the discovery of asymptotic freedom called for better experiments at higher energy, to test the detailed predictions about the violations of Bjorken scaling described by the Altarelli-Parisi equations \cite{AltarelliParisi1977}\cite{Dokshitzer:1977sg}\cite{Gribov:1972ri}. 

Large projects unavoidably entail long and international consultations and preparatory steps. The natural forum for these discussions was ECFA\footnote{European Committee for Future Accelerators.}. The politics was not restricted to a single laboratory. DESY, under the leadership of Herwig Schopper, had become a fully European Laboratory, competing with CERN for dimensions and scientific results. 

At the beginning of the 1980s, Schopper left DESY to become DG at CERN and Volker Soergel became director of DESY. Both laboratories had running machines: 
\begin{enumerate}
\item	DESY: PETRA was running since 1978 (until 1986) and the next step had to be discussed;
\item	CERN: SPS running since 1976 and S$p\bar p$S started 1981 and  the next step had to be discussed.
\end{enumerate}

It was an option for DESY to build an $e^+e^-$ machine near to Hamburg, since there was a long tradition of electron colliders at DESY, just as was for hadron accelerators at CERN. 

The decision eventually went at CERN in favour of (the future) LEP. Instead, an electron-proton machine was proposed (1983/4) for DESY, which became HERA under the leadership of Soergel and of Bjorn Wiik, who later, in 1993, succeeded Soergel as Director of DESY.\footnote{HERA (Hadron-Electron Ring Accelerator) was built in a tunnel located under the DESY site around 15 to 30 m underground with a circumference of 6.3 km.  Inside this tunnel, leptons and protons were stored in two independent storage rings on top of each other. HERA began operating in 1992, with four large experiments: H1, ZEUS, HERMES and HERA-B. Electrons or positrons were collided with protons at a centre of mass energy of 318 GeV. It was the only lepton-proton collider in the world while operating. HERA was closed down on 30 June 2007.}

The construction of the Large Electron-Positron collider at CERN started in September 1983, under the direction of Emilio Picasso, with Schopper Director General. The first collisions were observed on July 4th, 1989, and the first $Z^0$ on August 23 of the same year. At the time, the direction of CERN had passed to Carlo Rubbia.\footnote{LEP was located in a tunnel at about 100 m underground, with a circumference of around 27 km, located  near CERN, in the region between the Geneva Airport and the Jura mountain. 
Starting from  January 2001, LEP has been dismantled and the same tunnel adapted to host the Large Hadron Collider (LHC) and its four experiments, ALICE, ATLAS, CMS, LHCb, also located underground.}

\vskip 0.2 cm

L. B. \hspace{0.2 cm} Which were the main goals of LEP?

\vskip 0.2 cm

L. M. \hspace{0.2 cm} Initially, LEP was to be focused at the $Z^0$ mass (that is at a total center-of-mass energy of about $90$ GeV), to study the neutral mediator of the weak interactions discovered at the $Sp{\bar p}S$,~which was expected to be abundantly produced in electron-positron annihilation. 
This phase was called LEP I, to be followed by a second phase, LEP II, where the energy would be increased to around $200$~GeV. 

Aim of LEP II was to study the triple interactions $W^+ W^-\gamma$ and $W^+ W^-Z^0$, the latter being the salient aspect of the Yang-Mills interaction on which the unified electroweak theory was based~\cite{Glashow1961}\cite{Weinberg1967}\cite{Salam1968}. These interactions are crucial for the consistency of the unified theory and had never been seen before. 


The production of $W^+ W^-$ pairs, at LEP, goes via two independent channels
\begin{equation}
(a):~e^+ e^- \to \gamma \to W^+ W^-~{\rm and}~(b):~e^+ e^- \to Z^0 \to W^+ W^- \nonumber
\end{equation}
The first is a purely electromagnetic process, already predicted in QED, while the second is the genuine signal of the Yang-Mills nature of the electroweak interaction. At LEP, the two amplitudes could be distinguished and studied separately from the kinematic characteristics of the overall reaction. 
\vskip 0.2 cm

L. B. \hspace{0.2 cm} At that time, which were the reasons to go for higher energies after LEP?

\vskip 0.2 cm
L. M. \hspace{0.2 cm} The studies I just mentioned did not exhaust the possible verifications of the Standard Theory.

For one, besides the intermediate bosons, there was one crucial particle missing, the Brout-Englert-Higgs boson \cite{Englert1964}\cite{Higgs1964} responsible for the breaking of the electroweak symmetry, which is the source of the masses of quarks, leptons and of the intermediate bosons. 
The search and the consequent study of the Higgs boson\footnote{In his paper on electroweak unification, S. Weinberg noted the existence of this additional particle quoting the Higgs paper and coined the definition of \textit{Higgs boson}. The name has remained in the jargon of particle physics and will be used here for brevity. The correct attribution was recognised in the Nobel Prize 2012, attributed jointly to Francois Englert and Peter Higgs (sadly, Robert Brout had passed away in 2011).}  
in the full range of masses indicated by theory was one of the objectives --- actually the most pressing one --- of a high energy machine after LEP. 

In addition, there was the possibility of entirely new particles at higher mass, for which theoretical ideas were developed in the 1980s (see Box 1).

In the mid-1980s, the need of a higher energy collider seemed so pressing as  to lead Carlo Rubbia to suggest that perhaps one could dismiss LEP construction altogether and jump directly to make a Large Hadron Collider in the LEP tunnel. The idea met with a strong opposition in CERN and was fortunately dismissed by the CERN Management.

 \vskip 0.3 cm 
\begin{tcolorbox}[breakable, enhanced]

\footnotesize{
{\bf Box 1. The quest for Higher Energy after LEP}\\
\vskip 0.05 cm
The Standard Electroweak theory does not give a precise prediction of the Higgs boson mass but it only leads to an estimate of an upper bound of about $800$~GeV. Such high masses were definitely not in the mass range accessible to LEP I and LEP II, the latter being limited to reach a Higgs boson mass not much above $110$~GeV. The search of the Higgs boson in all its possible range of existence provided a first strong motivation for higher energy. Another was the search for higher mass quarks, such as the top quark.

An independent reason emerged in the early 1980s, when several theoretical papers pointed out the possible existence of particles beyond the Standard Theory, that could help to resolve the so-called \textit{hierarchy problem}. 

The essence of the problem is the extreme disparity of the mass scale of the Standard Theory, of the order of $10^2$~GeV as indicated by the $W$ or the Higgs boson masses, and the mass scale that would characterise a quantum theory of gravity, namely the mass obtained by combining  the Newton's constant,~$G_N$, characterising the strength of the gravitational interaction, with $\hbar$ and $c$ (Planck's constant and the velocity of light in vacuum). One finds the so-called Planck mass $M_P\sim 10^{19}$~GeV, about seventeen orders of magnitude larger than the electroweak mass scale. 

It was argued that quantum corrections to the Higgs boson mass are so singular in the Standard Theory as to bring its mass up to the Planck mass.

It was also observed that a new symmetry relating the scalar Higgs particle to spin 1/2 particles could suppress the quantum corrections so as to decouple the Higgs boson mass from the Planck mass. Such a symmetry would imply the existence of new particles in an estimated mass range of one to few TeV.\footnote{$1$~TeV=$1000$~GeV, about $1000$ times the rest mass of a proton.}

A symmetry of this kind, called \textit{Supersymmetry} had been mathematically characterised at the beginning of the 1970s~\cite{Wess1974}
 \cite{Akulov1974} and recognised to be an important step towards the unification of particle interactions, the Standard Theory, with gravity.

A different solution to the hierarchy problem was advanced in the 1980s, namely that, in analogy to the Cooper pairs in the theory of superconductivity, the Higgs boson is not elementary, but is rather a composite particle made of a fermion-antifermion pair, bound by new forces analogous to the colour forces that bind the quarks in the hadrons, which have been given the name of Technicolor forces \cite{Weinberg:1975gm}~\cite{Susskind:1978ms}.

The basic underlying theory would be based on elementary spinor and gauge fields only and, in this case, the conventional symmetries present for these particles would be enough to suppress the quantum fluctuations without having to resort to Supersymmetry. In this view, one finds again the prediction of numerous new particles in the TeV range, the bound states of the new fermions that make up the Higgs boson, called Techni-hadrons for brief.

In the same years, there was the confirmation by astronomical observations of the existence of a large amount of non-luminous mass around the galaxies, the \textit{dark mass}, that could not be made by atoms or by other particles of the Standard Theory, neutrinos included. The possibility that the lightest of the new particles required by Supersymmetry could be the constituents of the dark matter and could possibly be observed in high energy collisions added interest in the realisation of the LHC, and is still considered today as an important motivation to step up in energy in the future, should the LHC not find new particles in its energy range.

These considerations amply motivated the study of a proton-proton collider to be installed in the LEP tunnel at the end of the electron-positron program. Given the size of the tunnel and the technology conceivable at the time, the proton-proton collider could reach a  mass discovery potential of about $2$~TeV and was named {\it Large Hadron Collider}, LHC for brief. Hadron is the collective name given to the particles affected by the strong nuclear interactions. Here it denotes the proton, to distinguish this collider from the more familiar electron-positron colliders.


 }
 \end{tcolorbox} 
\vskip 0.3 cm

L. B. \hspace{0.2 cm} Before we move to the LHC, could you give a quick view on the LEP results?  Did LEP keep its promises?

\vskip 0.2 cm
L. M. \hspace{0.2 cm} Definitely yes. 
The LEP experiments produced a fantastic progress in our confidence in the Standard Theory.  

Of course, they have not been alone: experiments made with the Stanford Linear Collider at SLAC, HERA at DESY, Tevatron at FermiLab have also given great contributions. But, in my opinion, LEP results stand up for precision and for the variety of phenomena that they have been able to address.
\begin{figure}[ht]
\begin{center}
\includegraphics[scale=.55]{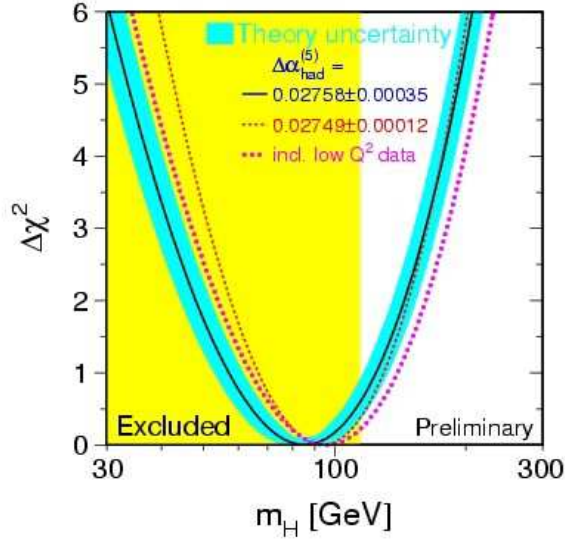}
\caption{{\footnotesize The Blueband Plot shows the constraints on the Higgs mass from precision measurements. The best fit and the width of the parabola vary, mostly due to shifts in the top mass and its uncertainty. In yellow, on the left, the region excluded by LEP experiments \cite{lepewwg}}.}
\label{bbplot}
\end{center}
\end{figure}

In extreme synthesis, four great LEP legacies are going to remain for long in our textbooks:\footnote{For a more exhaustive account of the LEP results see~\cite{Treille2002}.}
\begin{enumerate}
\item {\it Precision test of asymptotic freedom}, with the determination of the  behaviour of $\alpha_{strong}$ up to $200$~GeV. This result originated from the study of the hadronic jets coming out of the decay of the $Z^0$, which allowed to study gluon emission at considerably large value of the momentum transfer, which are controlled by $\alpha_{strong}$ and by the theoretical characteristics of the colour group. LEP tested QCD in the  region of large \textit{time-like} momenta. In the same years, the study of scaling violations at HERA produced values of  $\alpha_{strong}$ at large \textit{space-like} momenta. The agreement between the two classes of experiments is theoretically very significant.  
\vskip 0.4cm
\item {\it Three light neutrinos}, from the $Z^0$ width. We can measure the 
total $Z^0$ width, from the line-shape of the resonance, and individual widths, from the peak of the cross-section for each {\it visible channel.} Subtracting the visible from the total width one obtains the width in the \textit{invisible channels} (i.e. particles that do not interact in the detectors). Assuming invisible particles to be neutrinos and dividing by the decay width into one pair of neutrinos of a given type, provided by the Standard Model, one obtains the number of light, left-handed neutrinos in which $Z^0$ may decay. The number obtained is very close to three, confirming the Standard Model picture of three quark and lepton generations. This number provides today an important term of comparison with Cosmological observations, which are sensitive to the gravitational effects of heavy or sterile (i.e. not coupled to weak interactions) neutrinos which would not appear in $Z^0$ decay. At the moment, there is no discrepancy with three neutrino types, in the Laboratory or in the Sky.
\vskip 0.4cm
\item \textit{The $e^+ e^- \to WW$ cross section}, that definitely proved the existence of the $WWZ$ vertex, required by Yang and Mills. The predicted cross-section for $W$ pair production in the old Intermediate Vector Boson Theory, no unification and no Yang-Mills, has a quite more divergent behaviour with energy, than observed. The confirmation of electroweak unification with the Yang-Mills interaction is quite spectacular. 
\vskip 0.4cm 
\item \textit{Electroweak couplings and masses of quarks, leptons and intermediate bosons}. LEP I and LEP II produced many electroweak observables to be compared with higher order electroweak predictions. With the top quark discovered, in 1994, and the precision attained at LEP, one could afford to make an overall fit to the electroweak observables, with the Higgs boson mass, $m_H$, the only unknown variable. This led to the famous ``Blueband Plot'' (Fig.~\ref{bbplot}). Including results from all facilities,\footnote{There still remains some tension between SLC and LEP concerning the asymmetry of $b$ quark in $Z^0$ decay, see~\cite{lepewwg}.}
 the plot presents the $\chi^2$ of the fit versus the Higgs boson mass. It gave, at the beginning of 2000s, a very significant indication of a {\it light Higgs boson}, with the upper bound: $m_H \leq 200$~GeV to $90\%$ confidence level. 
\end{enumerate}

The consistency of the experimental results with the predictions of the theory was decisive for our confidence in the theory, and added interest to going at higher energy to see whether the Standard Theory had to be merged into an even more successful theory. Meanwhile, the search of the Higgs boson, within the LEP II mass boundaries, became the first objective of the experimental collaborations.





\section{The LHC Era Begins}

L. B. \hspace{0.2 cm} Back to the story of LHC. You mentioned Rubbia's idea of dismissing LEP in favour of a proton collider. How was it related with discussions about the choice  electron-positron  $vs.$  proton-proton machines?

\vskip 0.3 cm

L. M.  \hspace{0.2 cm}
Already before LEP was built, there was the idea of replacing LEP with a proton-proton collider in the  tunnel. 

The discussion started when the width of the LEP tunnel had to be determined. Most  Northern countries, in particular UK, wanted a tunnel diameter as small as possible, not to jeopardise LEP approval. My suspicion is that these countries also thought that LEP had to be the last CERN machine. But there was another party which argued that making a larger tunnel would make it possible, in a second time, to install a proton collider, a machine  with superconducting magnets appropriate to go to higher energy. The tunnel diameter was fundamental, because you need very intense magnetic fields to go to high energy, requiring extra space to install such magnets. And at the end there was a real confrontation in the LEP working group chaired by Nino Zichichi. 

The group had an advisory role  on the project made by CERN itself --- which was called the pink book. And they finally made a special recommendation to make the tunnel as large as possible in order to build later a superconducting proton machine. A compromise was reached, which at the beginning looked  very strange: to make the tunnel large enough to contain one proton superconducting ring, but not two. 

Unlike electron-positron, proton-proton colliders, in the most direct realisation, require two rings, because the charges of the colliding beams are the same, and therefore that seemed a contradictory measure. But it was accepted: it was better than nothing. In any case, the tunnel was made much  larger than needed for LEP and that stimulated CERN to think about the possibility of installing two proton beams in the same ring, what was called the {\it two-in-one} scheme. Two-in-one was not obvious since it required a magnetic field modulated so that it went one way in the position of one beam  and the other way in the position of the other. 

A two-in-one model was first designed  by Giorgio Brianti in 1985 and presented at the Moriond School. 
This may be
 considered to be the birth  of the LHC \cite{Brianti1985}.
 
 
 
The design presented  by Brianti had all the essentials of the LHC as realised later. The design was revised by Lyn Evans who made changes, ameliorations, changing the length of the magnets, but the essential ideas of Brianti's design have remained. 


\section{Colliders in competition}
Then came the Aachen meeting, in 1990, a workshop gathering some 500 physicists to put the research case for the proposed LHC. George Kalmus, in his closing remarks, declared: ``It (the Aachen meeting) has marked a watershed, the time, when the LHC project \dots \ graduated \dots \ to being the way forward for European particle physics.'' 
\begin{figure}[ht]
\begin{center}
\includegraphics[scale=.38]{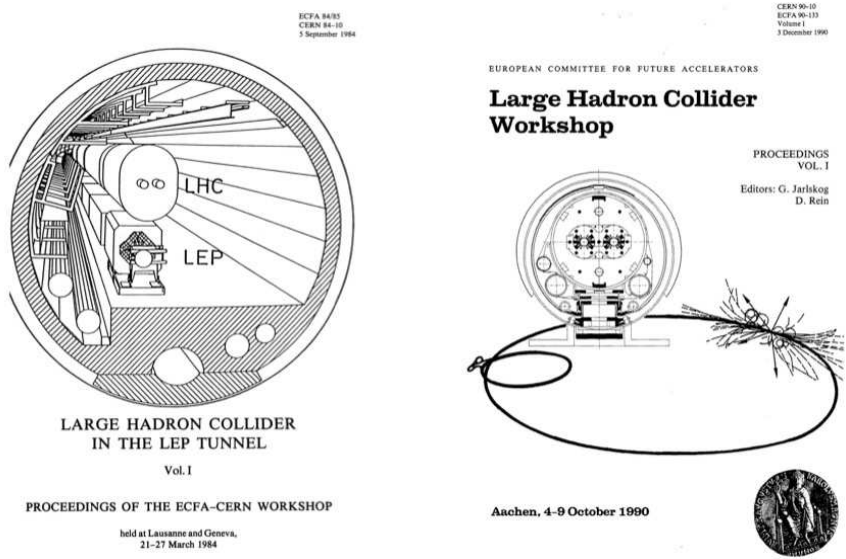}
\caption{{\footnotesize First pages of the Proceedings of the LHC workshops held in Lausanne, 1984  \cite{EcfaCern1984}, and Aachen, 1990 \cite{JarlskogRein1990}. Note the early design of the machine layout in the LEP tunnel, with the LHC and the LEP rings together. Coexistence of the two machines was considered up to 1995, when it was dismissed by Chris Llewellyn Smith at the Lepton-Photon Conference in Beijing.
}}
\label{cover}
\end{center}
\end{figure}

In 1992,  Council declared that the LHC ``will be CERN's next facility.''
In the Evian meeting, same year, Expressions of Interest for experiments were presented and the LHC Experiments Committee was created at CERN.  

My direct interest started in 1993 when, while President of INFN, I was nominated in the Italian delegation to the CERN Council and we started discussing how to proceed to secure the future of High Energy Physics in Europe. 

At the time, the Superconducting Super Collider (SSC) was being built in the US, in a new, very large tunnel which allowed to reach an energy quite higher than the LHC. There was a debate in Europe whether simply to drop CERN plans and join the Americans, or otherwise continue with the LHC project.

Somewhat before, Herwig Schopper, as CERN's director, had made part of a delegation to the US, to discuss a possible European participation in the SSC. In his book 
Schopper reports of a meeting in Washington where they were told that Europe should contribute between 100 and 200 million dollars \cite{Schopper2009}. But they were also told  that the project was there and they simply had to send the money. At these conditions, Schopper recalls, they could not accept, and the discussion was dropped there. 
It was a wise decision because in this way Europe did not commit itself to the complicated story of the SSC, which was totally out of our control since 
all decisions were made in the Senate of the United States. We avoided the fate of the Japanese, who committed themselves to the SSC and then learned from newspapers that it had been cancelled (late 1993).

Before that, there had been an important development due to Carlo Rubbia. He made a working group, with Sam Ting and others, to study the effect of luminosity on the discovery potential of an hadron collider. SSC was planned to have an energy of about 40 TeV, almost 3 times what LHC could reach, but designed to work at the typical luminosity of proton colliders, something like $10^{32} cm^{-2} s^{-1}$.  Carlo proposed what was called, a little emphatically, the {\it energy vs. luminosity trade}. 

In an hadron collider, protons have enormous energy (20 TeV in SSC, 7 TeV in LHC)  
~but you see the collisions of the most numerous partons, which have energies much lower that proton's energy. This makes, roughly speaking, a reduction factor of about 1/4 of the discovery potential with respect to the beam energy (assuming new particles are produced in pair). Thus a proton machine such as the SSC would discover new particles up to something like 5 TeV. 

If you make a machine with larger luminosity --- they were thinking of 2 orders of magnitude --- and you run it for, let's say, one year, you have a certain number of collisions, and among them there are also collisions of the partons which are more energetic than the average and therefore produce particles of higher mass than expected from the previous estimate. With a lower luminosity these collisions would not have sufficient statistics to be analysed. 

The result was that increasing the luminosity of the LHC by 2 orders magnitude, one could get to a potential for the discovery of new particles with mass of 2.5--3 TeV, which started being competitive with the SSC. The  {\it energy vs. luminosity trade} was used as an argument to convince the Europeans that it was still sensible to go for the LHC, given the obvious advantage of being a cheaper machine and easier to be done because the tunnel  infrastructure and the laboratory around it did already exist. 

In contrast, the SSC was planned to be an entirely new infrastructure, to be built in Texas from the green field, together with an entirely new laboratory. I think that it was in fact the big mistake of DOE not to make the SSC at FermiLab, but that is another story. 

The {\it energy vs. luminosity trade} argument was not entirely convincing. The SSC could have increased its luminosity in a second time, while the LHC could not easily increase its energy, but it kept most of European physicists behind the LHC, until finally the SSC was cancelled \cite{Riordan2000}.

\vskip 0.3 cm

L. B. \hspace{0.2 cm}
What was the reaction at the news that the Congress had cancelled the SSC project in  late 1993?

\vskip 0.3 cm
L. M.  \hspace{0.2 cm}
Well, CERN was ready to face the shock. However, it was a big shock. 

The argument  made by several countries was: If the Americans gave up why do we have to spend money to do this? So there was a big confrontation in the countries. 

Rubbia did not succeed in getting LHC approved by the CERN Council and the matter was left to the new Director General, Chris Llewellyn Smith. Chris took the approval of LHC as his main task. 

First, he appointed Lyn Evans as responsible of the machine, after Brianti retired, a really excellent move. Then they started a large R$\&$D effort on the superconducting magnets, which were the real issue. 

The other issue was the R$\&$D  on detectors, to make them able to stand  a luminosity never tried before of  $10^{34} cm^{-2} s^{-1}$. At CERN, Rubbia had made a committee to survey the issue, the LHC experiment committee, in which I have been in  1992-1993. Detector research was located essentially in CERN Member States, in Italy at INFN, and that created a large momentum in support of LHC approval.
\vskip 0.3 cm

L. B. \hspace{0.2 cm}
The high energy and luminosity essential for the physics goals of LHC gave rise to new formidable experimental challenges\dots \ In 1992,  letters of intent to build  the multipurpose detectors ATLAS and the Compact Muon Solenoid (CMS) were submitted,\footnote{ATLAS Collaboration, ``ATLAS. Letter of Intent for a General-Purpose pp Experiment at the Large Hadron Collider at CERN'' (CERN/LHCC/92-4, LHCC/I2, 1 October 1992); CMS Collaboration, ``The Compact Muon Solenoid. Letter of intent for a General Purpose Detector at the LHC'' (CERN-LHCC-92-3, LHCC/I-1, 1 October 1992).} 
 and in 1993  the ALICE Collaboration proposed to build a dedicated heavy-ion detector, expecting the formation of a new phase of matter, the quark-gluon plasma.\footnote{ALICE Collaboration, ``Letter of Intent for A Large Ion Collider Experiment at the CERN Large Hadron Collider'' (CERN/LHCC/93-16, LHCC/I4, 1 March 1993).} Technical proposals followed in 1994, when LHC was approved.

The search for the Higgs boson of course influenced the design of the ATLAS and CMS experiments. Which new physics scenarios were expected to be explored at LHC?
 
 
\vskip 0.3 cm

L. M.  \hspace{0.2 cm} The first requisite for the general purpose detectors, ATLAS and CMS, was of course to be able to discover the Higgs boson up to $800$ GeV, the upper bound indicated by the Standard Theory and also by possible variations such as the composite Higgs models (see Box 1). The signature of choice was identified in the decay $H\to \gamma \gamma$ and that led to the special requirement of detectors with a very high resolution in energy, to resolve a photon-photon line in a very large background of photons from $\pi^0$ decays. To this aim, ATLAS developed very sophisticated liquid argon detectors, CMS an advanced generation of scintillating crystals. In addition, there was the exploration of the high energy side, hunting for the hypothetical new particles. The key here was to develop very hermetic detectors, which could measure with great accuracy total energy  and total momentum of the particles produced in the collision. It was in fact estimated that a fraction of the new particles would decay in particles that would not interact in the detector, neutrinos or even the hypothetical dark matter particles. An anomalous signal of ``missing energy'' and ``missing momentum'' in the final states would be a clear signal of new physics. 
 
 The other detectors, LHCb and ALICE had more specific objectives. For LHCb: advanced resolution in energy and in the location of the vertex (to detect events in which weakly decaying particles were produced, such as  particles containing $b$ or $c$ quarks). For ALICE: the capacity to deal with events with exceptionally large multiplicity of final particles, such as those expected in heavy ion collisions.
 
 \section{Superconductivity in Italy}

L. B. \hspace{0.2 cm} 
LHC would use existing CERN infrastructure:  the LEP tunnel and the accelerator complex as an injector, which brought about a dramatic cost saving compared to creating a totally new facility. However, to reach the desired ultra-high beam energies, a very significant effort had to be undertaken to develop the large superconducting magnets. At that time, there was a program of cooperation between Italy and CERN on such an enterprise\dots

\vskip 0.3 cm
 L. M.  \hspace{0.2 cm}
Interest in Italy on superconducting magnets had started with the machine HERA, at DESY, when Italy decided to contribute not in cash, but providing a number of superconducting dipoles. It was an idea of Nino Zichichi, devised to prepare the Italian industry for the accelerators of the future and it was a very good intuition. 

The Italian companies interested were  Ansaldo Magneti, Zanon for cryogenics, and Europa Metalli  (a company owned by the Orlando family and later sold to the Outokumpu group in Finland) for superconducting cables. These companies had provided the HERA magnets in time. That was an encouraging success for Nino, who was thinking about the construction, in Italy, of a really big machine, Eloisatron (Eurasiatic Long Intersecting Storage Accelerator), a proton-proton collider of 100 TeV per beam \cite{Zichichi1990}. So he wanted to have the Italian industry to start into that business.

Nicola Cabibbo became INFN President after Nino and continued to promote superconductivity.  

A superconductive cyclotron was planned in Milano. At the time there was a very brilliant accelerator physicist in Milano, Francesco Resmini. INFN gave him the leadership for the  construction of the ``Ciclotrone Superconduttore'', that was completed by his collaborators after the untimely Resmini's death. The superconducting cyclotron was then transferred to Catania, where it is still working. 

After Resmini, the development of superconductivity was led by Lucio Rossi in LASA (Laboratorio di Superconduttivit\`a Applicata, in Milano) and that laboratory took over the collaboration with CERN on the LHC superconducting magnets. 

When I became President, succeeding Nicola, I strongly supported that collaboration, and extended the aim of the superconductivity program in Italy  to include the LHC detectors, for which superconductivity was becoming a central issue.

 We got extra money from the Ministry of Research to support R$\&$D on detectors in a joint, INFN-Industry collaboration. It was called the {\it 2\% project}, because it was based on money taken by the Ministry from the research Institutions in Italy, in the measure of 2\% of their funds, to be given to projects in collaboration with industry. The project with industry allowed us to recuperate money taken out of the INFN budget and put them to use for the future LHC experiments.

 The know-how thus acquired enabled later Ansaldo to win the contract for the cryogenics of CMS and LASA could participate in the making of one of the superconducting rings of ATLAS, in competition with Orsay. 

\section{The Approval of LHC}

L. B. \hspace{0.2 cm} When were the first superconducting magnets completed?

\vskip 0.3 cm
L. M.  \hspace{0.2 cm} The first 11 meters long magnet, made in collaboration between CERN, INFN and Ansaldo, was presented at CERN at the end of 1994. The magnet arrived just before  Council and it made a rather spectacular impact.
At the Council, Llewellyn Smith submitted again the proposal to approve LHC construction. 

At that point it was clear that Germany was resisting.  In part, I guess, because 
Bj\"orn Wick, then DESY director, did not believe that the two-in-one magnets would work. For this reason, the prototype was important, because it proved the two-in-one to be a viable concept. 
There were still questions about the optics, about joining one magnet to the other, but these were mechanical problems and the general idea was that they could be overcome.

However, there was still a problem of money, and many countries in the Council did not want to invest so much in the LHC. 

It was at this point the Chris and Lyn invented a way out which was called the ``missing magnet proposal''. The proposal was to approve as baseline a reduced project, which would contain only two thirds of the magnets and of course would cost less. In addition,  Council would give two years to the Management to find external contributions to fund the missing magnets and be able to submit to  Council the full project.

It was a brilliant move, which succeeded in removing the residual doubts. Many people thought that the missing magnet project could not be realised, because of the complicated cryogenic structure, so finding the extra money was really vital. 

But Chris accepted the risk and, in December 1994, the ``reduced'' LHC was approved.

\vskip 0.3 cm

L. B. \hspace{0.2 cm}
What was your personal opinion about all this?

\vskip 0.3 cm
L. M.  \hspace{0.2 cm}
 I was very happy for the physics. The approval of LHC meant we could continue doing first class particle physics for a couple of decades, at least.  Even though I felt, as many others, that the missing magnet solution could not stand and it was only a temporary solution.
 But I was happy also because we could go on with an important program in superconductivity to which I had already committed INFN.  
 
Chris and Lorenzo Fo\`a --- who was at the time CERN's Director of Research --- started a campaign to build consensus around the LHC. They opened discussions with many countries, in particular Japan, United States, Russia and India, and obtained a lot of support. The Japanese finally agreed to provide 300 million Swiss Francs, in money, suggesting that this money could be spent in Japan. As a consequence, Japan got orders for part of the superconducting cable. 

At the time, a very influential study appeared in the US  by a committee led by Sydney Drell. They observed that the plan of US financing for Particle Physics still contained a bump corresponding to the SSC expenses. The recommendation was to use this bump to finance a strong American participation in the LHC. For the American contribution to the machine, they could put to work the infrastructures made in Fermilab to build the superconducting magnets of ~SSC. Eventually, the US agreed to produce the LHC quadrupoles and other items, for a value of about 200 Million USD, a very important contribution, and the ``Drell bump'',  as it was called, was also used to finance the US participation to CMS and ATLAS. The total agreed contribution was  of 500 Million USD. 
 
 This was happening in 1994-1995. Then, in summer 1996 there was a new crisis of the cost. 
 
 Germany stated that, due to their difficulties with reunification, they wanted to reduce their contribution to  all research initiatives. In the case of CERN, the cut  envisaged was $8.5\%$ for the first two years and $9.3\%$ for the following years (as mentioned in \cite{Llewellyn2015}, see also~\cite{Schunck2016}). UK joined in, requiring a reduction in the LHC cost, and other countries, including Italy, stated that a reduction could only be general, for all Member States, and not allowed to Germany alone.
 
 There was a very complex discussion, between summer and the December Council, with the proposal of an overall cut to CERN's annual budget. On the whole, integrating over the years of LHC construction, the cut corresponded to about 600-700 million Swiss Francs. 
 
 Chris, at the end, accepted the cut, stating that it could be compensated by future savings from the laboratory. But there were energetic protests from the part of the CERN staff against Management, accused to have yielded to Council at the expenses of CERN's scientific future (see e.g.  \cite{MaianiBassoli2013}). 
 
  The Americans also were unhappy. In the CERN-US discussions it had been explicitly stated that the extra money given by the US would be used to improve the project and not for Member States to save on their contribution to CERN. 
 
 Finally, at the end of 1996,   Council approved (i) to proceed with the construction of the full LHC (which was good) but also (ii) to cut CERN budget (not so good). Everybody was happy and, essentially, everybody immediately forgot about the cut: somebody was  to take care of it later. 
 
In the same Council of December 1996, I was nominated President of the Council, to succeed the French scientist  Hubert Courien.
Courien had played a crucial role in the negotiations of the LHC with Germany and the other Member States, with vision, wisdom, political skill and determination.

 \vskip 0.3 cm
\begin{figure}[ht]
\begin{center}
\includegraphics[scale=.35]{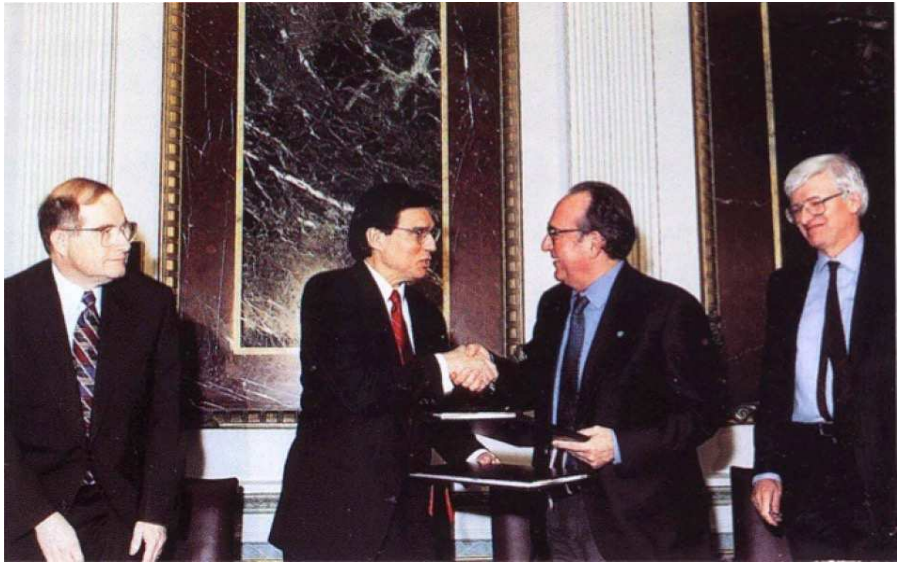}
\caption{{\footnotesize Signature of the agreement between CERN and the United States for participation to the LHC project. From left: Neil Lane, president of NSF, Federico Pe\~{n}a, Secretary of State, Luciano Maiani, President of CERN Council, Chris Llewellyn Smith, CERN Director General, Washington, December 8, 1997 (Photo CERN).}}
\label{IS}
\end{center}
\end{figure}

\section{On Board}

In 1997, the succession to Llewellyn Smith was opened. At mid year it appeared that I could be a suitable candidate. Chris  told me: it's either me or you. I suspended my function of President of  Council, and I  was proposed by the Committee of Council in September. 

However, formally, I  was still President of CERN Council in November '97, and I went to Washington with Chris  to sign the agreement with the US. The agreements with Japan and Russia had been already finalised, by Hubert Curien. The signature of the US agreement was my last act as Council's President and I was voted as Director General elect in December 1997, to take the direction of CERN one year later, starting from January 1, 1999. 

LHC construction, started in 1997, was now on my shoulders. 

At that time, it really seemed that LHC would go by itself. Lyn Evans was a well established project leader, as Picasso had been for LEP, there was a full team established, and the construction was going. 

In '99, with the LHC  we encountered only normal difficulties. One example arose in the excavation of the CMS shaft. CMS is very close to the Jura (much closer than ATLAS) and there is a real river flowing underground there, which made it impossible to excavate the shaft that had to bring the CMS detector to the cavern below. 

The solution was to inject liquid nitrogen in the ground, making a sort of permafrost in which it was possible to excavate, to return to normal conditions later, after the walls of the shaft had been consolidated. It was a brilliant solution, but it  did cost a considerable amount of time and of, not foreseen, money.

At that time, enough magnets had been produced to make a section of LHC --- a string of magnets 120 meters long --- which was installed in the laboratory to test the behaviour of the magnets, for example their quenching characteristics. The results of the tests were very positive, very informative and very encouraging. 
 \vskip 0.3 cm
\begin{figure}[ht]
\begin{center}
\includegraphics[scale=.45]{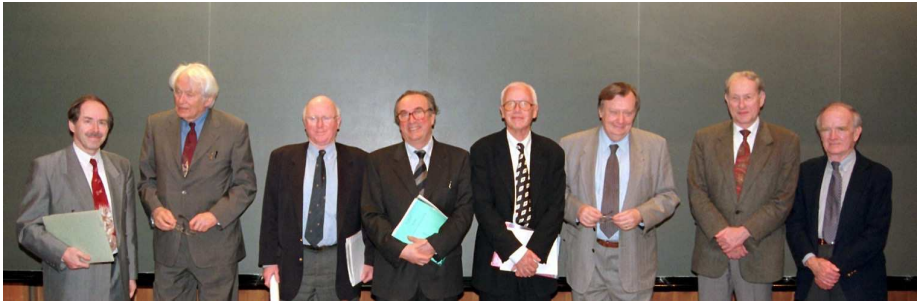}
\caption{{\footnotesize Seminar in honour of Carlo Rubbia on the occasion of his 65th birthday, 16 March 1999. Left to right: Gerard 't Hooft, Georges Charpak, Alan Astbury, Luciano Maiani, Klaus Winter, Carlo Rubbia, Arthur Kerman and Val Fitch (Photo CERN).}}
\label{IS}
\end{center}
\end{figure}
 So, I was taken by another story. 
 
 By 1998 neutrino oscillations of atmospheric muon neutrinos had been officially anounced~\cite{Kajita:1998bw}
 \cite{Fukuda1998}. However, only the disappearance of neutrinos had been proved. It was thought that $\nu_\mu$ would oscillate into $\nu_\tau$ but the low energy of atmospheric neutrinos did not allow to produce the $\tau$ in their subsequent interaction and to identify positively the $\nu_\mu - \nu_\tau$ oscillation. CERN could make a neutrino beam of sufficient energy, to be sent to the Gran Sasso laboratory in Italy, some 700 km south of Geneva. There, neutrinos would interact with a suitable detector, to be revealed as $\nu_\tau$ \cite{Elsener1998}. An Italy--Japan collaboration proposed to do so, with an ad hoc detector based on nuclear emulsions, later realised under the name of OPERA, led by Kimio Niwa and Paolo Strolin \cite{EreditatoEtAl1998}. 

In 1997, I had already expressed to Council my intention to realise a ``long-baseline neutrino beam'', going from CERN to Gran Sasso. Not everybody was happy but, after some discussion, the CNGS project (CERN Neutrinos to Gran Sasso) went through, with the external contributions from France, Germany and Belgium and the important support of INFN, which funded the decay tunnel at CERN and the preparation of the areas in the Gran Sasso underground laboratories to receive the neutrinos.
 
 \vskip 0.3 cm
 L. B. \hspace{0.2 cm}
At CERN the CHORUS experiment \cite{Eskut1997}, using the neutrino beam of the SPS accelerator, had completed data taking in 1998. What was the interest of performing such a new attempt if compared with other parallel efforts searching for a conclusive evidence of the oscillation mechanism at the atmospheric and solar scale? And especially in the context of  the new generation of long baseline experiments designed to search for a direct evidence of  $\nu_\mu - \nu_\tau$ oscillations?
 
  \vskip 0.3 cm
  L. M.  \hspace{0.2 cm}
 Neutrino experiments at CERN and other laboratories had searched without success  for the disappearance of muon neutrinos, but the parameters of this kind of oscillations derived later from the atmospheric neutrinos indicated that the oscillation length needed to reveal the effect was much longer and could only be obtained with a long baseline beam, like CNGS. At that time the only existing experiment of this kind was the K2K (the neutrino beam starting from the laboratory KEK located in Tsukuba, Japan, to reach the underground Kamioka Observatory, located in Kamioka, about 250 km away) which, however, had not the energy to produce the $\tau$s and therefore to prove positively the existence of $\nu_\mu - \nu_\tau$ oscillations. CERN had the right energy and Gran Sasso was at about the right distance. 
 
\begin{figure}[ht]
\begin{center}
\includegraphics[scale=.50]{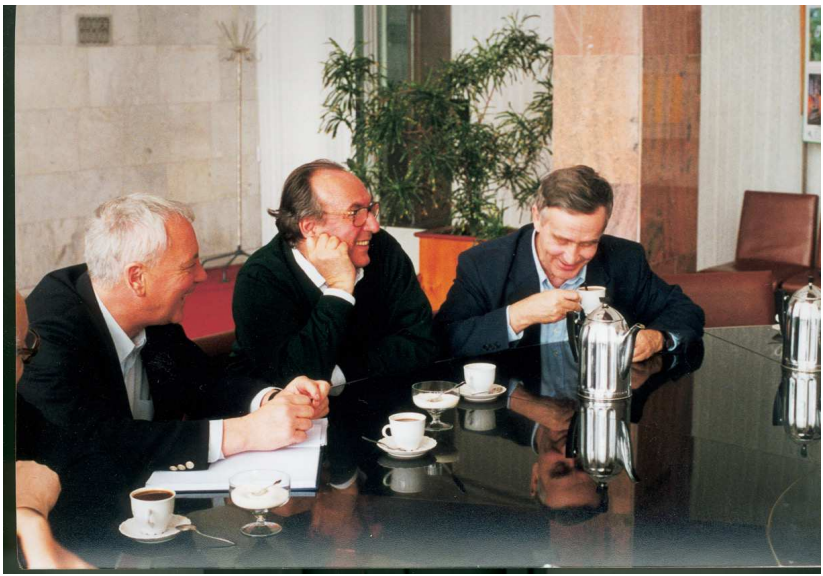}
\caption{{\footnotesize At the Budker Institute, Novosibirsk, to survey the production of Russian magnets, with Lyn Evans (left) and Sasha Skrinsky (right), in 2000. The warm magnets produced at Novosibirsk made part of the Russian contribution to the LHC and have been installed in the tunnels to transfer the proton beams from the SPS to the LHC ring.  Photo by A. Skrinsky.}}
\label{IS}
\end{center}
\end{figure}

\section{The End of LEP}

L. B. \hspace{0.2 cm}
In the meantime, what was happening with the tunnelling works for LHC?

\vskip 0.3 cm
L. M.  \hspace{0.2 cm}
In 2000, the excavation was getting near the LEP tunnel and at the end of 2000 LEP had to be discontinued. 

The original program was to shut down LEP at the end of 1999. But since excavation of LHC tunnels was late, Chris proposed to Council an extension to the full year 2000, which was approved just before Chris left, at the end of 1998. 

So, at the end of 1999 LEP entered into its last year. We were of course monitoring the results and around summer the ALEPH experiment collected some anomalous events that could indicate production of the Higgs boson, just at the upper end of the machine energy range. 

ALEPH reported three anomalous events with four jets in the final state. In the three events, two jets had originated from a $b$ quark (or antiquark) as indicated by the fact that the jets initiated at a distance from the point of the primary collision, i.e. had originated from a weakly decaying particle compatible with a $b$ mesons. The other two jets were compatible with the decay of a $Z^0$. The total mass of the $b$ jets clustered at the end of the phase space around $M=114$~GeV. In all, the events looked like:
\begin{equation}
e^+ e^- \to Z^0 + H
\end{equation}
followed by:
\begin{equation}
Z^0 \to 2~ jets;~H\to b {\bar b}
\end{equation}

Of course, these events had a background of similar events produced by interactions  not involving the Higgs boson, the distinction could be only statistical. At the end of summer, the significance of the three events to originate from the Higgs boson  was estimated by ALEPH to be $2.2~\sigma$ (standard deviations) the conventional limit for discovery being $5~\sigma$. 
The other experiments did not have anything so spectacular and reported only slight excesses of events over the no-Higgs expectation.

Roger Cashmore, was following them closely, and I was continuously in touch with  him during summer. 
At the end of summer, the experimental collaborations, led by ALEPH, asked to continue  for another year, to clarify the situation. 

Now that was a very big problem, because the time plan in the contracts of the LHC civil works established that excavations had to break into the LEP tunnel at the beginning of 2001. To give LEP the prolongation of one year, we should tell the companies to stop  for one year, but nonetheless continue to pay them all the same. Pending a more complete discussion, we prolonged LEP running, due to stop in September, to the end of October.

In fall 2000 there was a big celebration for the end of LEP running with authorities from Member States invited. There was happiness, but also apprehension about the pending closure of LEP. Steve Meyers, a little dramatically, called his talk the ``LEP wake''.

At that time, in view of the continuing expectation for new anomalous events, I asked discretely the LHC team to make a study about how much we should have to spend,  in case of one year prolongation. The answer was an extra cost of about 120 MCHF. 

Finally we arrived at the beginning of November which was the time when we should take the decision. In this discussion, I was in touch with Roger and with Michel Spiro, the Chair of the LEP experiments Committee. 

 \vskip 0.3 cm

L. B. \hspace{0.2 cm}
What did you think of the intriguing observation reported by the ALEPH experiment?

\vskip 0.3 cm
L. M.  \hspace{0.2 cm}
Well, we were not really convinced. But then, just at the end of October/beginning of November L3 recorded an event, which looked like something decaying to $b\bar{b}$ and produced in association with missing energy which could be a $Z_0$ going into 2 neutrinos. 

The picture was: you produce the Higgs and the $Z_0$, the Higgs goes into $b\bar b$ materialising into hadron jets and the $Z^0$  into $\nu-\bar{\nu}$. The mass of  the Higgs boson candidate was estimated at 114 GeV, similar to what had been found in the ALEPH events.

I was very much impressed and, at first sight, I thought that this event could change the entire world. But then, looking at it closer --- and Michel and Roger had an important say --- we realised that there were other ways but neutrinos from a $Z^0$, in which energy could be lost, which made the presence of a $Z^0$ not so clear. Neutrinos mean that energy is missing, but energy could be missing for other reasons: radiation going down the LEP pipes, or neutrinos from $b$ decays. Discovering $Z\to \nu \bar \nu$ had been possible at the $Sp{\bar p}S$ because both energy {\it and} momentum were missing. But the neutrino momentum, in L3, was related to the momentum of the $Z^0$, which essentially vanished because this event was at the top of LEP energy. No missing momentum was there to be seen and that meant that you would not be able to tell whether the missing energy was neutrinos or had some other explanation.

When the L3 event was presented in CERN --- by the physics coordinator, Simonetta Gentile, with Ting in the audience --- I had prepared one transparency to illustrate my point. 
\begin{figure}[ht]
\begin{center}
\includegraphics[scale=.60]{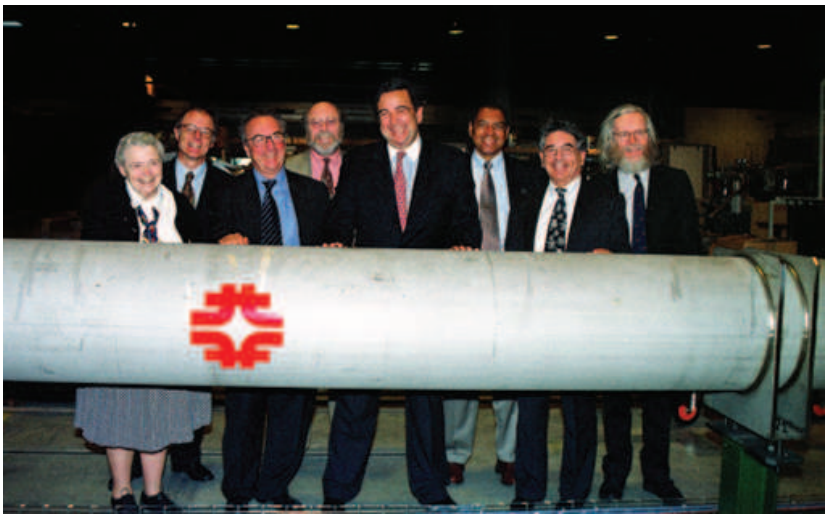}
\caption{{\footnotesize Mr. Bill Richardson, State Secretary of Energy of USA (center) behind the cryostat for a prototype superconducting  quadrupole magnet built by the US-LHC collaboration (hence the FermiLab logo) as part of the US contribution. Left to right: Mildred Dresselhaus, Carlo Wyss, Luciano Maiani, Roger Cashmore, Amb. George Moose, Peter Rosen (from the US Delegation at CERN Council), John Ellis. CERN, 16 September 2000 (Photo CERN).}}
\label{IS}
\end{center}
\end{figure}

``I am very sorry, I said, your interpretation may well be right, but the evidence is not conclusive.  This is not a golden event with neutrinos, because of the lack of missing momentum. It would be different if you could increase LEP energy by say 30 GeV. In that case you would produce the Higgs and the $Z^0$ with some momentum and neutrinos could be identified by missing momentum in addition to missing energy.  

In the present conditions, neutrinos can be identified only statistically, but in one year of running you will not be able to go from a suggestion to discovery. The significance of  the event as Higgs production was slightly below  $3 \sigma$, but to attain $5\sigma$, the discovery limit, would  require running for four to five years, or something like that.'' 

So I concluded: ``I am sorry but we cannot afford to go on.'' 

The committees that had to recommend the extension were not able to arrive to a decision. They had mixed opinions, some people were rightly afraid that a LEP prolongation might badly damage the LHC. Finally, the issue arrived at the level of the Director General and we had to take ourselves the decision. We made a directorate, we discussed around the table and arrived at the conclusion that it was not wise to put our money in such a risky enterprise as stopping LHC for one year. 

We left one door open: if Council would give us 120 million Swiss Francs to do that, we would take the risk.  But Council said no, you have to find the money yourself. Then I called for November 17 a special restricted Council Meeting (what was called the ``Committee of Council'') to prepare the decision for the December Council. The final CC statement read:

\textit{
On 17th November 2000, the CERN Committee of Council held a meeting to examine a proposal by the Director-General concerning the continuation of the existing CERN programme, which foresees the decommissioning of the LEP accelerator at the end of the year 2000.
The Committee has expressed its recognition and gratitude for the outstanding work done by the LEP accelerator and experimental teams.
It has taken note of the request by many members of the CERN Scientific Community to continue LEP running into 2001 and also noted the divided views expressed in the Scientific Committees consulted on this subject.
On the basis of these considerations and in the absence of a consensus to change the existing programme, the Committee of Council supports the Director-General in pursuing the existing CERN programme.
}

 \vskip 0.3 cm

L. B. \hspace{0.2 cm}
One can imagine that all this developed a lot of controversies about that decision\dots 

\vskip 0.3 cm
L. M.  \hspace{0.2 cm}
Yes! And for different reasons. The most obvious one was that there was  a whole community of people that really saw in LEP the last possibility to get the result of their life, I mean, for example, people that were going to retire and could not see a future in the LHC. Also, many people really believed that the anomalous events were the signal of a Higgs boson with mass $114$~GeV and that in one year they could get to a discovery.

\begin{figure}[ht]
\begin{center}
\includegraphics[scale=.50]{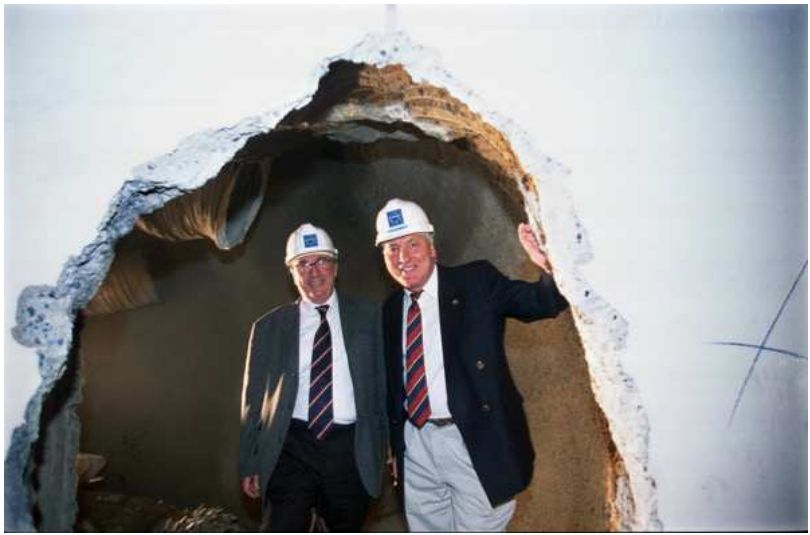}
\caption{{\footnotesize Getting into the LEP tunnel from the SPS with Lyn Evans (right), May 15, 2001 (Photo CERN).
}}
\label{cover}
\end{center}
\end{figure}

On the other side, LHC was still in an uncertain future. While the work of excavation was going forward, the preparation of the detectors was delayed. So, some people in the experimental collaborations thought that after all, waiting for one more year could not be a bad idea.

I was accused, by still another party, to have made a deal with the Americans selling them the Higgs boson. By stopping LEP, we were giving  Tevatron  a chance to discover the Higgs, in exchange of their support to continue with LHC. However, in the same meeting where we took the decision, Lyn Evans had made absolutely clear that Tevatron could not put LHC in danger, because of its too low luminosity. Indeed, LHC had suffered a delay of four years on the schedule we fixed in 2001, and still Fermilab could not see a competitive signal of the Higgs boson. Lyn's was a fairly accurate assessment.

So there was a lot of opposition, but we stood firmly by our decision. Very important was the support of Roger, Michel Spiro and Lyn Evans.

I presented the case in December and Council formally approved. 

But, even after Council, there were rumours that, after all,  LEP could start again. 

What made me jump was to learn, in the directorate of January, that the main power lines arriving to LEP had not been cut. So, I put in the minutes that if that line was still there in one week, I would go down in the tunnel and cut it myself.

The line was cut and in May 2001 the excavation of the tunnel connecting SPS to LHC broke into the LEP tunnel (Fig. \ref{cover}).

In spring, a more advanced analysis of the LEP events was presented. DELPHI and OPAL had not found any significant event. the analysis showed that the anomalous effect, combining the result of all collaborations, was of $2.3~\sigma$ (final results are reported in \cite{Barate2003sz}). It confirmed that, even if real, one year of more running could not possibly take the signal out of background. In fact, the most likely interpretation of the L3 event was that the two $b$'s had decayed semileptonically, with neutrinos taking out a lot of energy and almost no momentum.  

\begin{figure}[ht]
\begin{center}
\includegraphics[scale=.45]{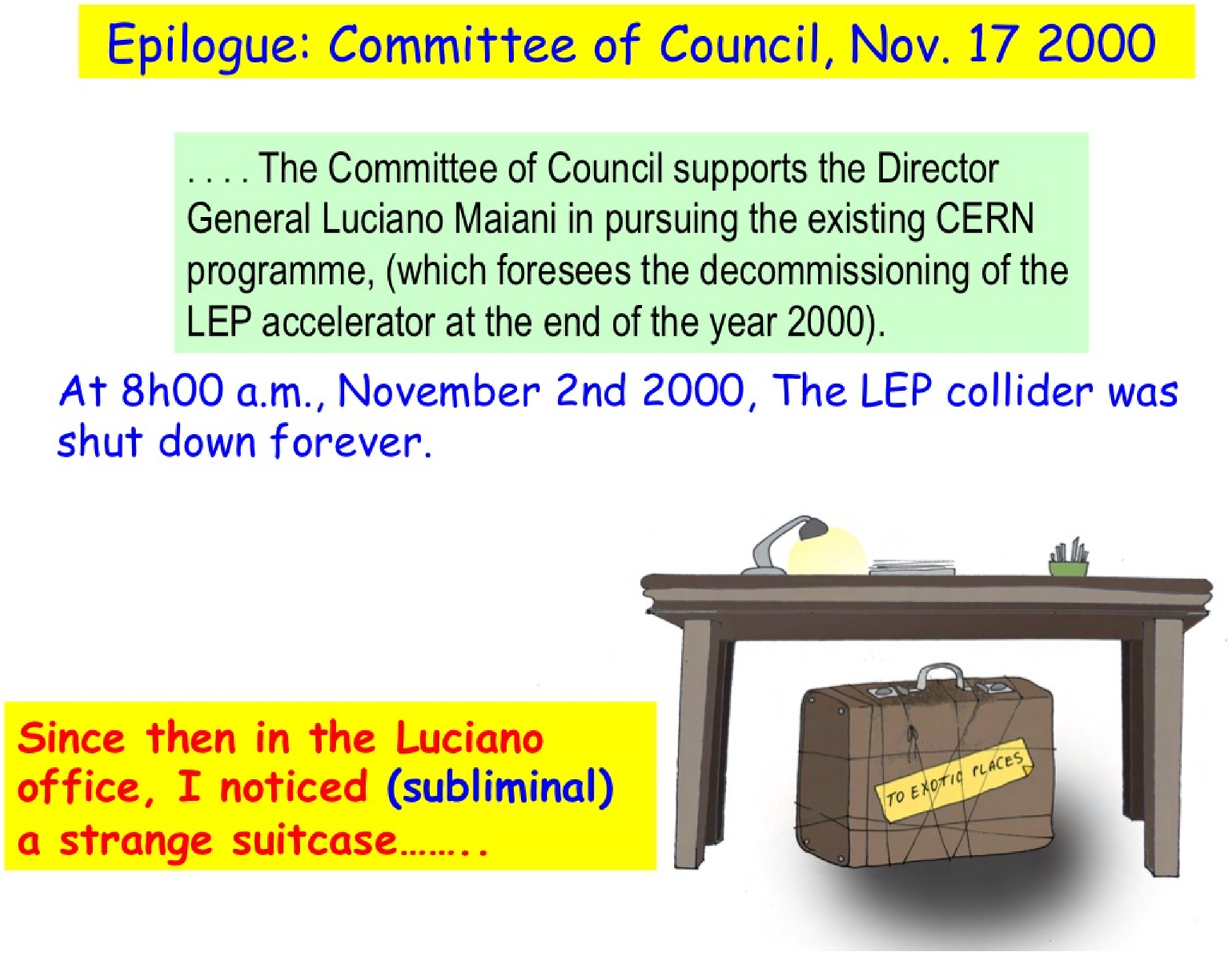}
\caption{{\footnotesize A slide presented by Carlo Dionisi at the Workshop for my seventieth anniversary, November 2011.
}}
\label{bag}
\end{center}
\end{figure}

Even after LHC went into operation, there were still people who believed that the Higgs boson was there in the LEP events. In 2011, during the celebration of my 70th birthday, a slide of Carlo Dionisi  (Fig.~\ref{bag}), showed my desk and below the desk a suitcase ready to leave for exotic destinations \dots \ just in case the Higgs boson were discovered at LHC, with a mass confirming the LEP events. 

Another post-LEP complaint was that Chris Llewellyn Smith had been offered by Ansaldo to buy an extra lot of superconducting cavities that could increase LEP energy by about 30 GeV. He had not accepted the offer for lack of funds. After that, Ansaldo dismounted the assembly line of the cavities and this possibility was not there at my time. Schopper, in his book quoted before \cite{Schopper2009}, criticises the choice attributing to the previous Director the fact that LEP had to be prematurely closed. The issue was raised again in a meeting of the Scientific Policy Committee that took place in 2012 with the participation of former members (the old boys). Chris was present and recalled that at the time he had not the money and that it was totally out of the question to ask Council to increase the CERN's budget for that. So, he concluded, I could not obtain the money and I did not have the money, and that was it. 

I must say here that I agree with Chris. If you have a program approved for a new machine, LHC, with an enormous financial effort, you cannot press Council because you want to spend an additional 15 million Swiss Francs to potentiate a machine arrived at the end of its energy, LEP, just to have another ambitious little territory.

It would have been different if CERN did not have any other plan. But CERN {\it was} committing itself to the LHC, the only machine capable to sweep in one year the whole Higgs boson region, which is what LHC did in the years 2011-2012.

In my report to Council in December 2000, about the decision to close LEP, I concluded:

{\it This decision moves us definitely into the LHC era. 
A powerful complex of machine and detectors, to fully explore the Higgs and SUSY region. Le Roi est mort. Vive le Roi!!}

How much we needed to focus all efforts at CERN on the new objective became clear nine months later, in fall 2001.  

\section{Cost Crisis}

\vskip0.2cm
L. B. \hspace{0.2 cm}
Well, this brings us to September 2001 and the disclosure of the cost increase of the LHC\dots
\vskip 0.3 cm

L. M.  \hspace{0.2 cm}
In 1999, we had made an exploratory call for tender for a first lot of superconducting magnets and the reply was very bad, I remember a cost about twice what had been planned. Lyn advised not to take the offer. He thought the companies were charging  us for the risks of construction, still poorly known. With more R\&D on cold mass assembly and more prototyping, companies would be more confident and would lower the cost. We did so, putting more R\&D resources into the magnets.

In spring 2001 we went to the call for tender for the full lot of  magnets. Offers came to CERN during summer. While in vacation with my wife (in Ischia) I remember being in touch almost  every day with Lyn Evans, to monitor the situation. We got a cost considerably less than what we had gotten in 1999, but still  a good 30\% more than expected. This time we had to accept.

So, when I came back in September, I asked to make an overall estimate of the total cost the LHC. I thought it was time to have not only the cost of the machine, but also to make an overall recognition of what would be the cost of everything was needed. The need of much larger computing power was already coming out, estimated at 150 MCHF and also the increase in CERN contribution to detector construction (50 MCHF). And there were the costs of installation, the upgrade of the injector system,  waste disposal, etc. In short, all what was needed to refurbish and upgrade CERN infrastructures, to host and run the LHC. 

These items had not been really ever costed, they had gone always under the entry: general laboratory expenses. I wanted to put everything together and find out what really was the ``cost-to-completion'' of the LHC and see how it would fit in the Laboratory budget of the coming years. 

We went through a very deep examination. There were joint meetings of the groups that were building the LHC, each of them reporting what had been spent and what they needed to complete their task. I was there all the time, carefully listening. 

Because in fact there was criticism. The previous attitude, when the cost had to be given at the time of approval, was to make what could be called ``planning for success''. If you tell me you need to spend 100,  I tell you that you have to get to 80 and put 80 on my list. And, of course, to get to 80 becomes your problem. 

I remember I was trying to do that again, but there was no way. This time, people did not want to sign for 80. So I got convinced that the extra cost was there. 

Besides the LHC machine and the new civil construction cost (the caverns for the experiments), there were all the other issues I just mentioned, never considered before. It came also out that Lyn Evans, with Chris authorisation, had spent 120 million in prototyping that had been added to the LHC cost, while, according to the plan made at the time, prototyping should have been charged to the Laboratory, to be reabsorbed within 2009 (another 20 Millions that I had authorised in 1999, at the time of the first call for tender of the magnets, had taken the same destination).  I was rather upset. ``You could have told me before --- was my comment --- fortunately they have been well spent \dots''.

So we put everything together. There was about 480 MCHF increase in the cost of the machine, corresponding to  $18\%$  increase over planned cost, and the total increase was estimated at 850 MCHF.  

To give the full picture, we thought it useful to add also expenses which we deemed necessary in the future, such as R\&D for a future linear collider or for the LHC upgrade, which brought the total above one billion Francs.

\begin{figure}[ht]
\begin{center}
\includegraphics[scale=0.35]{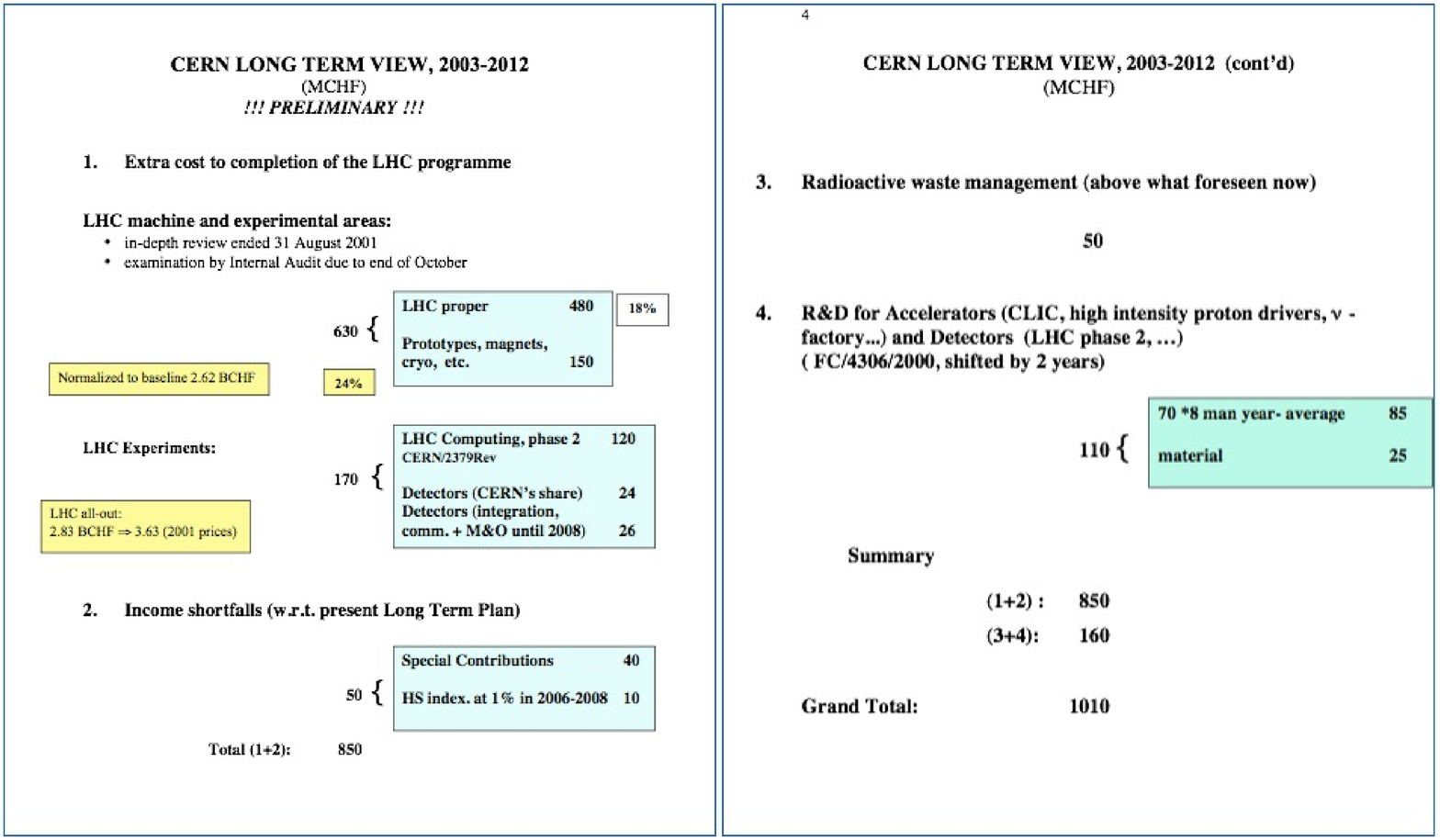}
\caption{{\footnotesize  
Items in the first page,  adding to 850 MCHF, are the extra cost to completion of the full LHC project, including additional infrastructures, computing and detector expenses, as reported to the Finance Committee of September 19, 2001. Items in the second page provide estimate of future expenses.}}
\label{crisis1}
\end{center}
\end{figure}

I wrapped up everything in two slides (Fig.~\ref{crisis1}), and called for a special session of the Finance Committee, on September 19,  to tell them our findings. 

In the Finance Committee, the 1 billion figure was like an atomic explosion. The Netherland delegate asked me to resign immediately, the French were very upset. Other delegations, however, were more in favour to see what we could propose. 

We needed to react quickly. Two days after the Finance Committee, we made the first decisions. To recuperate the prototyping money out of a reduction, smeared over the years up to 2009, in CERN expenses on R\&D, except for the Compact Linear Collider (CLIC), and in Lab consolidation. This would reduce LHC overcosts by 180 MCHF.

 To assess how much could be further recuperated, we needed a fast and detailed review of all Laboratory activities. To this aim, we appointed five ``Task Forces'' chaired by recognised experts from the Lab on five critical issues: Research Programs, chair Dieter Schlatter, TF1;  Organisation, chair Horst Wenninger, TF2; Industrial Support and Contracts, chair Karl Heinz Kissler, TF3; Personnel, chair  John Ferguson, TF4; LHC and other Accelerators, chair Steve Myers, TF5. The Task Forces started immediately and produced first results before the Committee of Council meeting of November.

We also scheduled a campaign of in-house communication involving myself and all members of the Directorate, to illustrate the situation with Council and our lines of action.

In the following days, we also realised that there was a fault of perspective in our way to see things, and a fault of communication  in our relations with Council.

At least unconsciously, we tended to blame Council for the 1996 cut. They had taken the water out of the laboratory in which the fish of cost overruns could swim (as it had always happened for large projects everywhere) and now it was up to them to put back the money. 

It was obviously a wrong attitude. Council delegates had followed the instructions of their Governments and had done their best to smooth out the difficulties to CERN. In addition, CERN had anyway committed itself to do the LHC within {\it that} budget, and now it had to comply --- a   conversation with Charles Kleiber, the Swiss Secretary  of State for Science, and a confidential message by Peter Minkowski, a distinguished Bern physicist who was advising the Swiss delegation, had been quite illuminating, in this respect.

Finally, putting all difficulties in a single package had been a clear communication mistake. 

To correct the communication side, I followed the suggestion of the Director of Administration, Jan van der Boone, and we asked a Public Relation company, Saatchi $\&$ Saatchi, to advise us in our communications to Council, starting with the Committee of Council meeting in November. It proved to be a very good idea and the support of an expert in communication has been crucial to overcome diffidences and misunderstandings.

The Finance Committee met on November 6. The expert of Saatchi $\&$ Saatchi had made long, soul searching, interviews with me and with the other Directors, asking about our assessment of the situation and our objectives. On the basis of a draft made by the expert, I prepared a written intervention to be read as an introduction to the financial situation and reported in the minutes of the meeting.

I started from the 1996 cut. The cut in the budget should have produced --- I admitted --- a thorough revision of CERN activities. But the urgency of the message has not been clearly perceived. 
The problem had been further complicated by the need to maintain the vitality of CERN and, in fact, high level scientific results had been obtained, such as the discovery of direct CP violation and the likely observation of a new phase of matter in heavy ion collisions. 

\textit {The way in which we at CERN addressed the LHC funding issue has not reflected well on us.   I want to apologise for this and make it clear that I accept full responsibility.} 

\textit{There have been failures in management  judgement and in communication --- especially  communication, I believe. But I'd like to stress that the LHC project itself remains technically and scientifically sound.} 

Then followed the key messages. 

{\it CERN was committed to deliver LHC within the agreed budget and it remains committed to doing this --- if Council so decides.  

My personal view is that steps necessary to deliver without additional funding may not be in best long-term interests of the Organisation.  Hence, we will be asking Council to consider other options. 

We recognise that none of these --- about to be presented in detail --- may be wholly satisfactory. Consider them as a starting point in a dialogue process.}

\begin{figure}[ht]
\begin{center}
\includegraphics[scale=0.45]{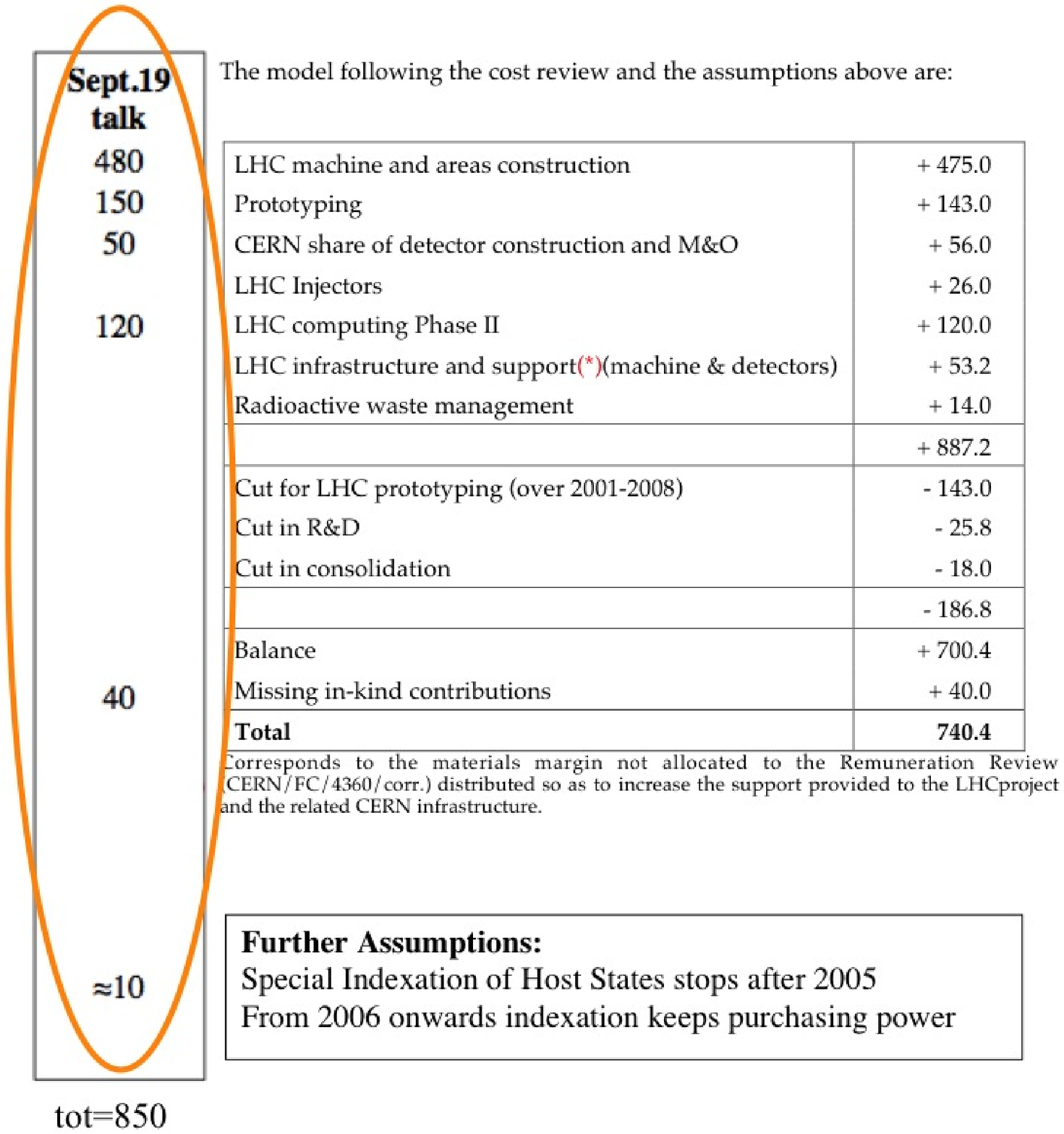}
\caption{{\footnotesize 
Extracosts of the LHC project as reported in the Finance Committee of November 6, 2001. The figures confirmed, with small variations, those reported to the Finance Committee of Sept. 19, shown in the column on the left.}}
\label{crisis2}
\end{center}
\end{figure}

In the following, I confirmed the September costs, informed about the first measures taken to absorb part of the deficit in the Laboratory's budget (Fig.~\ref{crisis2}), and gave details on the Task Forces that had to determine, within a short delay, the extent to which the Laboratory could absorb the extra deficit and show where we would {\it hit the hard rock} (i.e. cuts would affect the LHC itself).

Another argument on the table, discussed at long by Finance Committee  delegates, was that of establishing an External Review Committee (ERC), to have an independent assessment of the LHC cost and of the ways the Laboratory was dealing with LHC construction. 

I was completely confident of our cost estimate, but got also convinced that an external review was a necessary step to re-establish Council confidence in the Laboratory and to support delegates action vis-a-vis their Governments, even at the risk of putting Management under external control. 
 
\begin{figure}[ht]
\begin{center}
\includegraphics[scale=.45]{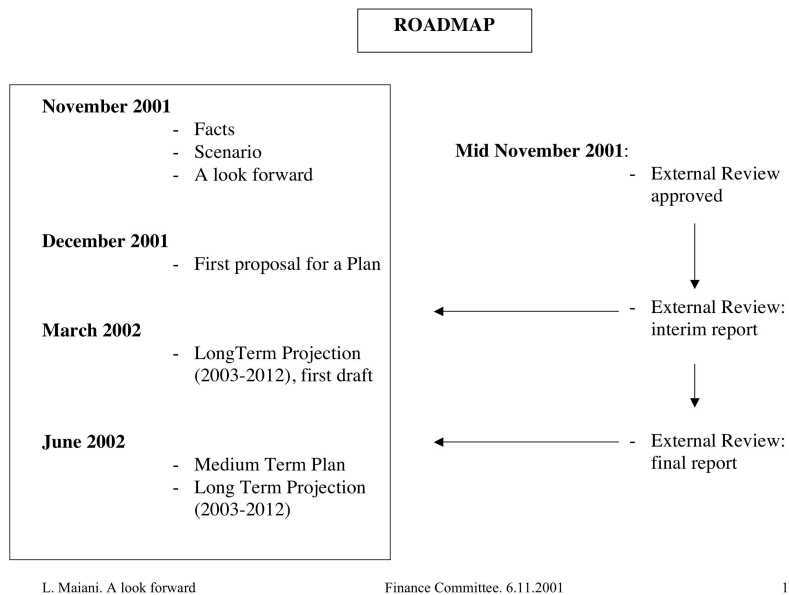}
\caption{{\footnotesize  
Slide presented at Finance Committee of November 6, 2001, with the roadmap to arrive in June 2002 to a new Medium Term Plan and a Long Term Projection including the revised LHC cost-to-completion.}}
\label{road}
\end{center}
\end{figure}

At the end, to give substance to our statements, I presented a roadmap (Fig.~\ref{road}), of the steps to include the new cost to completion in a Medium Term Plan and a Long Term Projection replacing the ones made before the crisis, in June 2001.  

In a letter sent personally to all members of staff,  I summarised the situation after the Finance Committee and tried to indicate how we could get out of the mess.

{\it Dear Colleague,

As you know the LHC will cost more than was originally planned.  CERN must do its part, and a sizeable fraction of the deficit will have to be found within our own budget. 

The time has come to examine how to re-align the future organisation of our Laboratory to this new reality. 

I am determined to keep you fully informed of developments.  All the facts are already on the CERN home page (see Special Announcements) but with this letter I would like to address you personally.

There is need for change. Resources must be focused on the LHC, in particular human resources.  
I am fully committed to leading this change. 

At the same time, it will be essential to preserve a limited but vital core of diversified activities and know-how, which make the strength of CERN and represent the seeds for the future of the Organisation and of particle physics, beyond the LHC.

The Finance Committee meeting on November 6 was the first step in the dialogue with the Member States. The discussion was critical but constructive with all delegations expressing strong support for the LHC project.

I have created special Task Forces to determine: the cost of different programmes, areas of savings across the Laboratory; restructuring; improved tools for managing CERN resources. 

Full information on the composition of these Task Forces is also available on the homepage.  I encourage you all to be active in this process. Your ideas are important.

All of us are committed to CERN as an institution and the ideals it represents. Doing what is best for science will be our guiding principle in moving forward. 

We have enormous challenges ahead and there are no short cuts. 

With your input and support I am confident we will define a solution, which will allow us to complete the LHC successfully and guarantee a future of scientific excellence for CERN.
}
\vskip0.1cm

\begin{figure}[ht]
\begin{center}
\includegraphics[scale=.45]{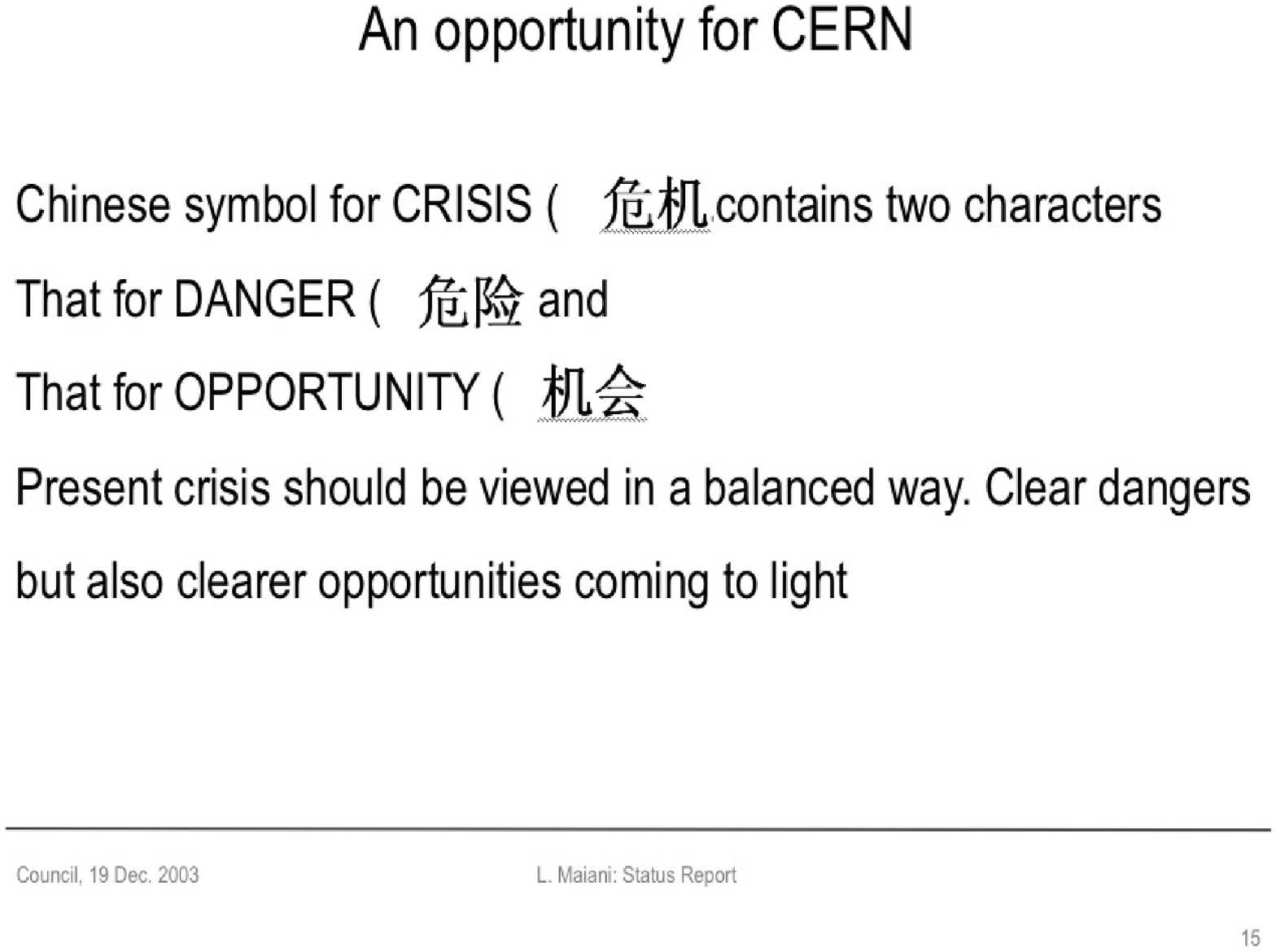}
\caption{{\footnotesize  Crisis = Danger + Opportunity, the motto of the years 2001-2003.}}
\label{opportunity}
\end{center}
\end{figure}

The motto of these days was that the crisis represented also an opportunity to strengthen structure and scopes of the Laboratory. This was exemplified in a slide I showed several times, passed to me by Jan van der Boone (Fig~\ref{opportunity}). A posteriori, I can say it is exactly what happened.

Since October 2001, I was working on what we would call later {\it a balanced package}. With the help of Task Forces, we could focus the Laboratory and simplify the Organisation, directing to the LHC the resources made available, in material and personnel. 
We could go roughly half way on the extra cost but the rest should be made available by Council. How?

I was already getting indications from Lyn that cable production was not going to meet the stringent time scales set in the contracts. Reporting  the commissioning of the LHC, one could recuperate the rest of the missing resources out of the margin that a constant budget contained in the extra years, without having to ask for an additional contribution from Member States, by then totally unrealistic. This is where the agreement of Council was essential and what was realistic to ask.

For the time being, however, this was  only a working hypothesis, vaguely alluded to in our communications. By the way, it gave substance to a mantra that Chris had been often repeating: {\it time is LHC contingency}.

The Committee of Council that followed took the new step, establishing an External Review Committee.

Following our proposal, the Committee of Council appointed Robert Aymar to chair the External Review Committee. A well known expert in superconducting magnets, Aymar had made a revision of the ITER project, including a new cost assessment, and was already familiar with the LHC, having refereed the project of the superconducting magnets.  The Committee of Council gave him the task of selecting his fellows commissioners, to be appointed by the Council of December 2001. 

The {\it balanced package} took shape in January 2001, with the first conclusions of the Task Forces just made available. 

In the Directorate, we made a review of superconducting cable production, with the conclusion that it could not be completed before the end of 2005. This meant reporting LHC commissioning to mid 2007, a two year delay with respect to the original 2005 schedule announced by Chris in 1996. 

Roger Cashmore insisted for a really in-depth examination, in regard to the experimental collaborations, which were not going to be happy with the delay, but Lyn did not leave any doubt. Even with the recent addition of a new company, Outokumpo from Finland, and with the new production centre in Brugg, Switzerland, the end of 2005 was just barely possible.

Reporting the LHC made it possible to follow one of the Task Forces suggestions, namely to close CERN accelerators for one full year, in 2005. We had already decided a 30$\%$ cut to the PS and SPS running time, but a one year suspension allowed to recuperate substantial material and personnel resources to focus on LHC installation. 

In addition, we reduced R\&D efforts in CERN to the development of the Compact Linear Collider (CLIC) only. Finally, the extension of the construction period allowed to attenuate the computing power problem, diluting the expenses over six, rather than four, years. 

With the experimental and accelerator programs thus restricted, the full experimental community insisting on CERN was aligned on the LHC program and no further cuts were possible, we had gone to {\it hit the hard rock}. We could now ask Council to extend by two years the constant budget regime agreed in 1996, from 2009 to 2011, and this would produce enough resources to complete safely the LHC. 

\begin{figure}[ht]
\begin{center}
\includegraphics[scale=.45]{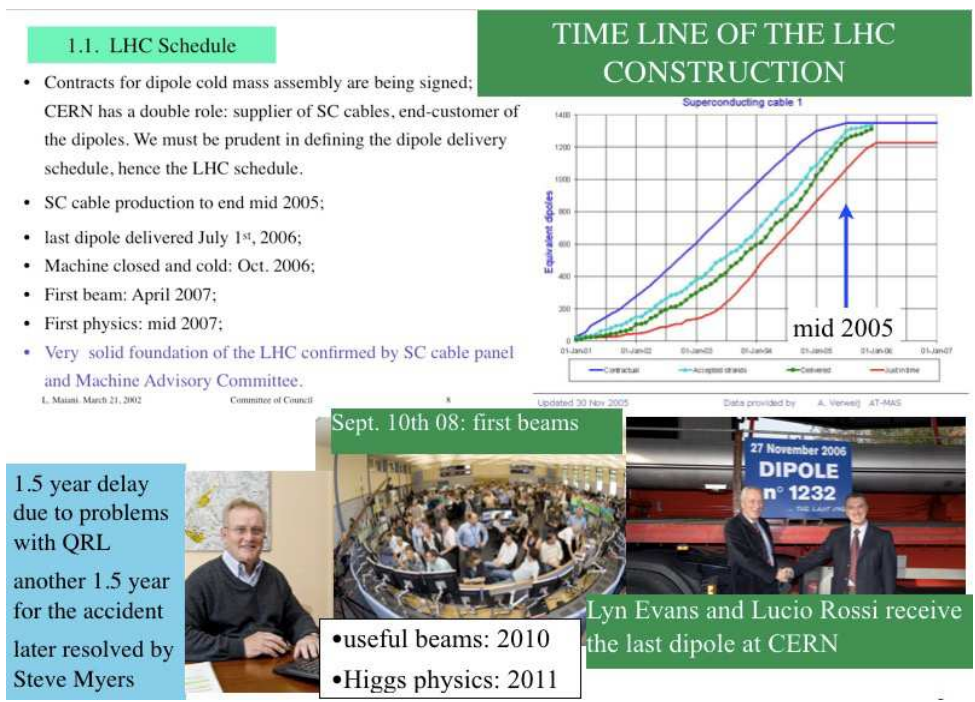}
\caption{{\footnotesize  The LHC timeline 2002-2012. Upper left: Schedule proposed in 2002. In clockwise order: Delivery of superconducting cables, completed in November 2005; Reception of the last superconducting dipole by Lyn Evans (left) and Lucio Rossi (right), November 2006; Beam commissioning, September 2008; Steve Myers, Project Leader from 2008; Delays due to QRL and to the 2008 accident; First physics 2011-2012.
}}
\label{timeline}
\end{center}
\end{figure}

We presented the new schedule to the Committee of Council in March 2002 (see Fig.~\ref{timeline}), and the new Medium Term Plan and Long Term Projection to the Council of June 2002, as promised.

 In spring 2002, another problem came to the fore. Companies want to be paid shortly after delivery and, with cables and magnets arriving in the Laboratory, payments would pile up quickly. We had a constant budget but were heading toward a peak of the spending. 

Of course, LEP had gone through the same problem, and it  was solved by taking a loan from a consortium of banks. We went back to the same consortium, but the answer was very negative.  At best, we would obtain only very expensive loans.
\begin{figure}[htb]
\begin{center}
\includegraphics[scale=.45]{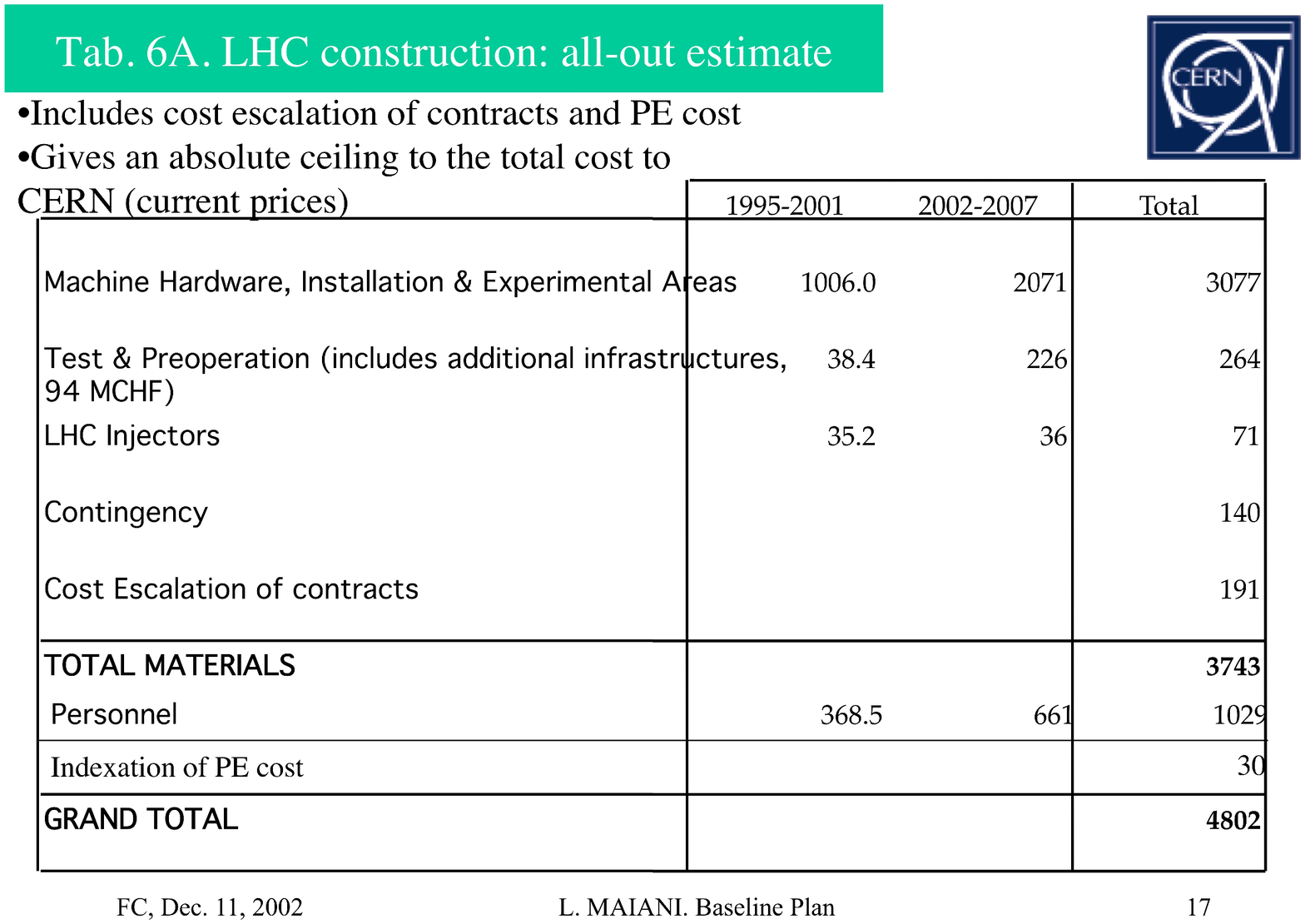}
\caption{\label{cost02} \footnotesize Estimate of the cost to completion of LHC and related infrastructures, in Millions CHF, presented at Finance Committee and Council of December 2002. For the final costs, see Tab.~\ref{costs08}.}
\end{center}
\end{figure}

A year before, in May 2001, we had created  an association with ESA, ESO and other European Research Organisations, called EIROforum (European Intergovernmental Research Organisations) and we had periodic meetings in Bruxelles with Philippe Busquin, the European Commissioner. During an informal dinner following one of these meetings, I asked Busquin: ``How is Europe going to finance the expenditure peak of ITER?''. Answer: ``The European Investment Bank will make a loan.''

So, back in CERN I sent Andr\'e Naudi, Head of the Finance Division, to discuss with the European Investment Bank (EIB) the perspective of a loan and it turned out that they were very positive. 

\begin{figure}[htb]
\begin{center}\label{signatureEIB}
\includegraphics[scale=.55]{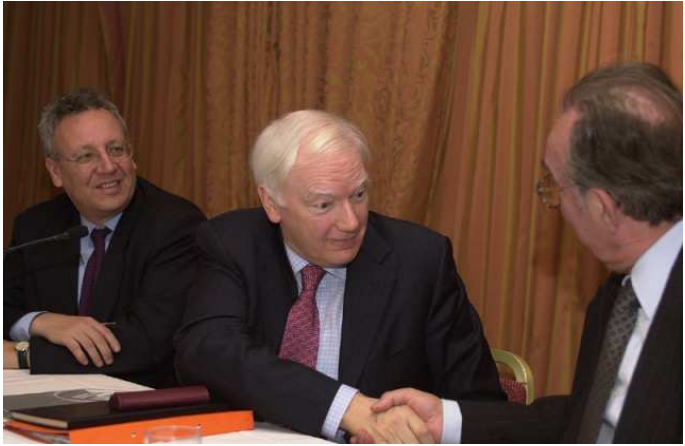}
\caption{\label{cost02} \footnotesize Signing of CERN's 300 million EUR loan from the European Investment Bank in December 2002. From left: Philippe Busquin, European Research Commissioner, Philippe Maystadt, EIB President, and Luciano Maiani, Director General of CERN (Photo CERN). 
}
\label{EIB}
\end{center}
\end{figure}

In December 2002, together with the approval of the new Plan, Council authorised us to finalise the negotiations with EIB, which ended up in Luxembourg (December 19, 2002) with the signature of a 300 million Euros loan, at $4\%$ annual interest rate (CERN got a triple A status) (Fig.~\ref{EIB}).  This was the final step for the resolution of the cost crisis started in September 2001.

The cost to completion of LHC and related infrastructures presented in December 2002, the best estimate that our Management could offer to Council, is reported on Fig.~\ref{cost02}.  Later, I was happy to see the agreement with the final result, reported in a CERN document in 2009 (Tab.~\ref{costs08}).

\vskip0.1cm
 \begin{table}[htb]

\centering
\small
\begin{tabular}{|l||rrr|}
\hline
                  &    Personnel     & Material    &Total \\
\hline
  {\bf Machine and Experimental Areas} &   {\bf 1 150}& {\bf  3 685} &  {\bf  4 835}  \\
  \hline
{\bf Injectors} & {\bf 86} &  {\bf 67} &  {\bf 153} \\
\hline
  {\bf Detectors: construction, R$\&$D} & {\bf 879} &   {\bf 312}  &  {\bf 1 191} \\
\hline
{\bf Detectors: test and pre-operation }& --  &  {\bf 181} &  {\bf 181} \\
\hline
{\bf LHC Computing}& {\bf 86} &  {\bf 93} &  {\bf 179} \\
\hline
{\bf Grand Total} & {\bf 2 202} &  {\bf 4 337} &  {\bf 6 539} \\
\hline
\end{tabular}

\caption{{\footnotesize Cost  to CERN of LHC and associated detectors, in Millions CHF. Source: CERN/2840, May 27, 2009.}}
\label{costs08}
\end{table}

A few final considerations about the works of the External Review Committee and the reactions of CERN personnel. 

The ERC was for us a powerful stimulus to go quickly to the solution of the cost problem, using our knowledge of the system and the results of the Task Forces. Indeed, we arrived at the March 2002 meeting with a plan already well developed, when the ERC presented the first report on their assessment of the situation. The good relations of the  Directorate with the ERC, Aymar in particular, did avoid dangerous confrontations. 
In substance, the ERC was functional not much to suggest the way out to the problem, as to certify for Council that our proposals were sound.

CERN staff was, like us, very conscious that the issue at stake was the very survival of CERN. With the communication line established by the Task Forces, tensions between Management and Staff were limited to a minimum, unlike what had happened  in the LEP story. 

All in all, I think the way the various actors behaved, Council, Management, ERC and Staff, makes a success story about how one can bring back to safety the trajectory of a large body running toward disaster, without making dramatic revolutions.

\section{The LHC Final Timeline}

 L. B. \hspace{0.2 cm}
From 2002 to 2012, there were still 10 years to go: what happened then? 
\vskip 0.3 cm

 L. M.  \hspace{0.2 cm}
Let me make a step back.

At the beginning of 2002, when Council started to look for the next Director General, I declared that I did not want to be prolonged in any case, because of my family problems. Pucci's very serious illness had manifested itself few months before, and developed during the time of the cost crisis until her death, in May 2003. 

At the end of 2002, Robert Aymar was elected Director General, to take function one year later, January 1st, 2004. 

It was a natural and very good choice, given his experience with large projects and the knowledge of the Laboratory he had acquired with the ERC. Working well with Lyn as Project leader, Aymar led the LHC to its first important success, beam commissioning, September 10, 2008, about 1 and 1/2 year later than foreseen in our 2002 schedule. 

In fact, the delivery of the superconducting cables had gone essentially on schedule and so the delivery of the dipoles, the last one arriving on November 27, 2006 (Fig.~\ref{timeline}), only five months off schedule.

What had gone wrong was the superfluid helium distribution line, that had been discovered at delivery not to comply with project specifications.  Finally, the distribution line was made at CERN, and that produced one good year delay. 

The rest is a well known story. 

One week after commissioning, a catastrophic superconductivity quench was caused by a faulty connection between two dipoles. In the following explosion, about fifty magnets were destroyed in the tunnel, fortunately with no human losses. 
The following analysis showed that faulty connections were spread all over the machine. 

Led by Steve Myers, who had replaced Lyn after his retirement, the CERN team introduced very sensitive resistance monitors, that would signal tiny losses of conductivity in time to prevent the quench, and made possible to run the machine at smaller energy and luminosity, starting at the beginning of 2010, with physics runs taking place in 2011.

After little more than one year running, as anticipated at LEP time,
 ~the discovery of the Higgs boson was announced in the CERN auditorium. It was a memorable session, chaired by the CERN Director General Rolph Heuer in the morning of July 4th, 2012, when Fabiola Gianotti (ATLAS leader) and Joe Incandela (CMS leader) announced that the signal observed by each experimental Collaboration had gone above the fateful level of $5~\sigma$s.

Francois Englert and Peter Higgs were both present at the session, which was video transmitted all over the world. Global Press cover was assured, well beyond the scientific interest,  by the (totally baseless) claims that the LHC collisions could produce a black hole that would eat up our whole planet.

Starting from spring 2013, magnet interconnections have been redone, Fabiola Gianotti has taken over as CERN DG since January 1st, 2016.  The LHC is now running under her leadership at full project energy and luminosity, searching for new phenomena above the Electroweak scale. 


The LHC story is very telling about modern particle physics and its people. 

The Higgs boson discovery is, of course, a great success of machine builders and tells a lot about effectiveness and continuity in CERN's Management. From LHC conception to the Higgs boson it took three Project Directors (Giorgio Brianti, Lyn Evans, Steve Myers) and five Directors General (Carlo Rubbia, Chris Llewellyn Smith, Luciano Maiani, Robert Aymar, Rolph Heuer) each carrying on his part of the burden during his term.  

The experimental Collaborations have also fared exceptionally well. The bet on luminosity has been completely won by detector builders, who are now planning a further increase in luminosity of a factor of ten, leading to the High Luminosity LHC. 

In the announcement of July 4th 2012, Fabiola Gianotti mentioned the great challenge of distinguishing among the several collision vertices from which particles originate when proton bunches cross, it was about 25 vertices/bunch crossing, but is now faring well above 100 vertices/ bunch crossing. 

She also mentioned the exceptionally big computing challenge posed by the analysis of LHC data, now well above Petabyte/year. 
The computing challenge was met with the LHC Data GRID: a network covering all the Globe, with primary nodes (Tier 1) in different continents where LHC data are stored and can be accessed by a community of users similarly spread all over the Planet.  

The Data Grid was proposed at the end of the 1990s ~\cite{Foster1999} and pioneered by CERN from 2002  onwards and by other scientific enterprises such as the Laser Interferometer Gravitational Wave Observatory (LIGO), and the Sloan Digital Sky Survey (SDSS). In the outside world, the concept is often referred to as The Cloud. After the World Wide Web, it is a major contribution of Fundamental Research  to our society.

\section{Future?}
\vskip 0.3 cm

 L. B. \hspace{0.2 cm}
After a forty-year search, the detection of the Higgs boson at CERN marked the beginning of a new era in particle physics, making Europe the undisputed leader in the field. But this is not the end of the story\dots What are the perspectives and new challenges of high-energy physics beyond LHC and the open questions that the LHC's successor should answer? Moreover, there is still the hot pending question of detecting dark matter particles\dots
\vskip 0.3 cm
\vskip 0.3 cm

 L. M.  \hspace{0.2 cm} The Higgs boson discovery is not the end of particle physics. 
 
 For a satisfactory theory of the fundamental interactions, there remain, among others, the problems of the mass hierarchy, the unification of  strong and electroweak forces and that of all interactions, including gravity. There is, in addition the problem to identify the nature of the ``dark matter'', revealed by astronomical observations to be the dominant component of matter in the Universe, perhaps related to the existence of the very same particles necessary to solve the mass hierarchy problem (see  Box 1).
 
 We may have guessed some real issues: compositeness, supersymmetry, but there are so many things we do not fully understand  that the physics we may expect to find beyond the Standard Theory is likely to be \textit{entirely new, strange and unexpected}. We need experiments to guide our intuition. 
 
 The key to the unification problems may well be at super-high energies, out of reach of today conceivable accelerator technologies and rather   
to be found with underground laboratories (e.g. the lifetime of the proton) and the observation of the very Early Universe. However there is the concrete  possibility that the answer for the hierarchy and the dark matter puzzles is resolved by particles with masses in the few to several TeV region (see again  Box 1), which are amenable to accelerator searches. 

A first exploration of this region will be performed with the ongoing Run 2 of LHC and, later with the High Luminosity LHC, an increase of  luminosity by a factor 10, which will bring the discovery potential of new particles around 3 TeV. 

It is also possible to double the LHC energy by utilising, in the same tunnel, more powerful magnets, based on Nb3Tin superconducting cables. It is hoped that in this way one could detect, in the next decade, concrete indications of new physics beyond the Standard Theory.

With the tunnel limitation, however, it is not likely that the LHC can get to a complete view of the new physics implied by, e.g. SUSY or Technicolor: we need to plan for more powerful accelerators. What's next?

An early proposal is the {\it International Linear Collider}, a linear accelerator producing $e^+e^-$ collisions at centre-of-mass energy of $0.5$~TeV. Purpose of the ILC would be to study with great precision the properties of the Higgs boson and search for deviations from the Standard Theory that would indicate new particles at higher energy, like the Blue Band plot (Fig.~\ref{bbplot}) did for the Higgs boson. 

With the sequence: LHC-ILC one would repeat the alternation of discovery/precision machines that was so successful with the S$p\bar p$-LEP sequence.

For the  ILC, there is one site approved in Japan (at Kitakami) and one reserve site (at Sefuri), but funding is still to be defined.

Alternatively, one may try to repeat the LEP-LHC strategy with a large circular tunnel. The proposal is to install in a large tunnel of 
 $70-100$ km a circular $e^+e^-$  machine with $250-300$  GeV  for precision study of the Higgs boson, to be replaced later by a $p-p$ collider of  $80-100$ TeV, to explore the full region left by LHC, from $3$ to $10$ TeV.

Two projects of this kind are being proposed. At CERN the {\it Future Circular Collider}, and in China the {\it Circular electron-positron Collider}, followed by the {\it Circular proton-proton Collider}, proposed by the International Institute of High Energy Physics in Beijing.  

A $100$ TeV proton Collider would be a fantastic challenge for new innovative technologies: material science, low temperatures, electronics, computing, big data. It would be a strong attractor of young talents and of new ideas, to solve the hardest scientific problems which we have been confronted in the last $100$ years.

The problem is, of course, that of resources, human and financial, for such big enterprises.

In the years 1950s, following the vision proposed in 1949 by the eminent physicist Luis De Broglie, governments of Italy, France, United Kingdom, Germany, Denmark etc., united forces to make CERN: the European facility that none of them could afford and which made it possible the exploration of the Microcosm that we have seen.

My hope is that, by the 2030s or so, regional Laboratories of Europe, America, Asia may unite in a Global Accelerator Network , that will make it possible to address the problems raised by the success of the Standard Theory and the discovery of the Higgs boson, by establishing and running the $100$ TeV World machine we dream of today.

\section{Acknowledgments}

I am grateful to Drs. Horst Wenninger and Hermann Schunck for a useful exchange about the LHC cost crisis. Constructive comments by the referee that helped making the story more accurate and complete are gratefully acknowledged.

\end{document}